\documentclass[10pt,twocolumn,aps,pra,groupedaddress,superscriptaddress,amsmath,amssymb,showpacs,english]{revtex4-2}

\usepackage{graphicx}
\usepackage{dcolumn}
\usepackage{bm}
\usepackage[T1]{fontenc}
\usepackage{textcomp}
\usepackage{mathrsfs}
\usepackage{amsmath}
\usepackage{bbm}
\usepackage[normalem]{ulem}
\usepackage{amsbsy}
\usepackage{braket}
\usepackage{amstext}
\usepackage{amssymb}
\usepackage{physics}
\usepackage{setspace}
\usepackage{esint}
\usepackage{xcolor,colortbl}
\usepackage[colorlinks,citecolor=blue,urlcolor=blue,linkcolor=blue,hypertexnames=true]{hyperref} 
\usepackage{babel}
\usepackage{booktabs}
\usepackage{mciteplus}
\begin{document}

\title{Quantum synchronization in XX chains under relaxation}

\author{K\"{u}bra S\"{u}mer}
\affiliation{Faculty of Engineering and Natural Sciences, Sabanci University, Tuzla, Istanbul 34956, Turkey}

\author{Ege Yiğit Aldırmaz}
\affiliation{Faculty of Engineering and Natural Sciences, Sabanci University, Tuzla, Istanbul 34956, Turkey}

\author{Steve Campbell}
\affiliation{School of Physics, University College Dublin, Belfield Dublin 4, Ireland}
\affiliation{Centre for Quantum Engineering, Science, and Technology, University College Dublin, Belfield, Dublin 4, Ireland}
\affiliation{Rinn Quantum National Research Centre, Ireland}

\author{Bar\i\c{s} \c{C}akmak}
\email{cakmakb@farmingdale.edu}
\affiliation{Department of Physics, Farmingdale State College—SUNY, Farmingdale, NY 11735, USA}

\author{G\"{o}ktu\u{g} Karpat}
\email{goktug.karpat@sabanciuniv.edu}
\affiliation{Faculty of Engineering and Natural Sciences, Sabanci University, Tuzla, Istanbul 34956, Turkey}

\date{\today}

\begin{abstract}

We study stable synchronization in qubit XX chains with site-dependent frequencies and exchange couplings under multi-local amplitude damping noise, extending the uniform-chain results of B.~\c{C}akmak \textit{et al}., [\href{https://doi.org/10.1103/yrfl-y3z8}{Phys. Rev. A \textbf{114}, L030202 \textcolor{black}{(}2026\textcolor{black}{)}}] in several directions. We establish that the dark-mode energies are precisely the eigenvalues shared by all undamped segments separated by the damped sites. Generic stable synchronization of the edge qubits occurs if and only if these segments share exactly one eigenvalue and this eigenvalue is nonzero, in which case every qubit on the support of the corresponding dark mode is synchronized with locking constants independent of the initial preparation. This spectral matching condition identifies the Hamiltonian parameters that support synchronization, while the couplings to the damped sites control the relative oscillation amplitudes. We present analytical examples which show how inhomogeneity can create synchronization absent in uniform chains and turn preparation-dependent locking into generic synchronization. We also calculate the asymptotic pairwise concurrence and prove that, in the single-dark-mode regime, every pair participating in synchronized oscillations remains entangled with a nonzero, constant concurrence.

\end{abstract}

\maketitle

\section{Introduction}

The phenomenon of synchronization occurs when oscillating systems adjust their rhythms, either to an external drive or to each other, and oscillate with a common frequency and a fixed phase difference~\cite{Blekhman1988,Pikovsky2001,Balanov2009}. It is observed in many natural and engineered systems, such as fireflies flashing in synchrony~\cite{Buck1968}, metronomes placed on a common movable base~\cite{Pantaleone2002}, and generators in power grids~\cite{Filatrella2008}. Studies include coupled oscillators inspired by cardiac pacemakers~\cite{Mirollo1990}, the Kuramoto model of phase oscillators~\cite{Acebron2005}, chaotic systems~\cite{Boccaletti2002}, and complex networks~\cite{Strogatz2001}.

Synchronization has also been studied in quantum systems~\cite{Schmolke2026,Solanki2026}, with several approaches developed to characterize the locking of their dynamics~\cite{Mari2013,Li2017,Galve2017,Eshaqi-sani2020,Li2022}. Early studies considered periodically driven dissipative two-level systems and qubits coupled to driven resonators~\cite{Goychuk2006,Zhirov2008,Zhirov2009}. Oscillators have since provided a common setting for studying both synchronization to a drive and mutual synchronization, including van der Pol oscillators~\cite{Lee2013,Walter2014,Lee2014,Walter2015,Sonar2018}, optomechanical systems~\cite{Heinrich2011,Ludwig2013,Qiao2020,Kalit2021}, and trapped-ion phonon lasers~\cite{Hush2015}. Alongside these studies, basic questions were raised about what constitutes a quantum limit cycle and how synchronization should be understood in systems with only a few levels~\cite{Roulet2018a,Parra-Lopez2020}. Questions about its characterization also arise in synchronization blockade, where the roles of symmetries, interference, and the synchronization measure have been examined~\cite{Solanki2023}, and synchronization was shown to pass through blockaded spins~\cite{Kehrer2024}. For larger systems, the discussion extends to collective synchronization and the influence of interactions, geometry, and topology on which parts of the system synchronize~\cite{Nadolny2023,Nadolny2025,Solanki2024,Wachtler2023,Wachtler2024,Liu2025,Wachtler2026}. Controlling the approach to synchronization and understanding its relation to quantum thermodynamics provide further directions~\cite{Impens2023,Murtadho2023,Razzoli2024}.

In open quantum systems~\cite{BreuerPetruccione}, dissipation can induce synchronization between undriven subsystems through a separation of decay rates. When a slowly decaying oscillatory mode of the Liouvillian dominates several local observables after a transient time, their oscillations acquire a common frequency and fixed relative phases~\cite{Giorgi2019}. This synchronization is transient when the surviving contribution eventually decays. Environment-induced synchronization has been investigated in harmonic oscillator networks~\cite{Giorgi2012,Manzano2013}, spin systems coupled to a common environment~\cite{Orth2010,Giorgi2013,Karpat2020}, and atomic lattices with inhomogeneous local losses~\cite{Cabot2019}. In dissipative spin systems, connections with subradiance and Liouvillian exceptional points have also been established~\cite{Bellomo2017,Cattaneo2021,Cabot2021}. Related transient synchronization can occur in closed harmonic networks without external dissipation~\cite{Benedetti2016}. Further studies have examined synchronization via repeated system-environment interactions and the influence of environmental memory~\cite{Karpat2019,Karpat2021}, including memory effects in driven systems~\cite{Xiao2023}. Dissipation-induced synchronization can also provide information about the spectral density of an environment~\cite{Giorgi2016}. Although synchronized dynamics can accompany quantum correlations~\cite{Giorgi2012,Manzano2013,Giorgi2013,Galve2017}, the formation of correlations between the synchronizing subsystems is not always necessary~\cite{Karpat2019}.

Stable synchronization requires oscillatory contributions which survive at long times. For time-independent Lindblad dynamics, these correspond to nonzero purely imaginary Liouvillian eigenvalues, whose relation to dynamical symmetries and synchronization has been established~\cite{Buca2019,Tindall2020,Buca2022}. A decoherence-free subspace (DFS) provides a mechanism for sustaining such oscillations, since the evolution within it remains unitary despite the coupling to the environment~\cite{Zanardi1997,Lidar1998,Knill2000,Lidar2003}. Related studies have considered decoherence-free normal modes~\cite{Manzano2013a}, nonlocal dissipation~\cite{Li2023}, phase locking in two-qubit systems~\cite{Sterba2023}, and persistent oscillations in collision models~\cite{Guarnieri2022,Li2023a,Campbell2025}. In spin chains, local white noise has been shown to produce stable synchronization between edge magnetizations~\cite{Schmolke2022}, while continuous measurement can induce synchronization along individual quantum trajectories~\cite{Schmolke2024}. Experiments have demonstrated synchronization to external fields in spin-1 atoms~\cite{Laskar2020}, nuclear spins~\cite{Krithika2022}, and trapped ions~\cite{Zhang2023,Li2025}, as well as mutual synchronization between two trapped-ion oscillators~\cite{Wachtler2025}. Quantum synchronization has also been simulated on IBM quantum processors~\cite{Koppenhofer2020}, and noise-induced synchronization of the edge qubit oscillations has been recently observed in a superconducting transmon chain~\cite{Tao2025}.

In a recent Letter, we established the conditions for stable synchronization in a uniform XX chain under local relaxation~\cite{Cakmak2026}. It has been shown that generic stable edge synchronization occurs if and only if a single one-excitation dark mode survives, with the same locking constant for every initial state producing persistent oscillations. The dark mode count follows from a greatest-common-divisor condition involving the chain length and damping locations, while spatial symmetry of the model fixes the edge oscillations to have equal amplitudes. However, exact uniformity can be difficult to achieve experimentally, and variations in qubit frequencies and couplings modify the mode energies and spatial profiles. 

In this work, we extend the uniform XX chain results to chains with arbitrary site-dependent frequencies and exchange couplings under local or multi-local relaxation. We characterize the full DFS through its one-excitation dark modes and establish that the unique dark mode condition for generic stable edge synchronization remains valid without spatial symmetry, provided the dark mode has nonzero energy. We then determine when this condition is satisfied by showing that the dark mode energies are precisely the eigenvalues shared by all undamped segments separated by the damped sites. This criterion gives necessary and sufficient conditions on the Hamiltonian parameters and recovers the greatest-common-divisor rule of the uniform XX chain as a special case. It also separates the parameters that determine the existence and energies of the dark modes from those that only shape their spatial profiles. Through  analytical examples, we show how inhomogeneity can lead to synchronization that is absent in uniform chains, turn initial state dependent locking into generic synchronization, and tune the relative amplitudes of the surviving oscillations. Finally, we extend the entanglement analysis beyond uniform chains by deriving the asymptotic concurrence for any pair of qubits. In the single dark mode regime, every pair participating in synchronized oscillations remains entangled with a constant concurrence.

This paper is organized as follows. In Sec.~\ref{sec:model}, we introduce the model. Sections~\ref{sec:dfs}-\ref{sec:sync} characterize the DFS, the asymptotic dynamics, and the conditions for stable synchronization. In Sec.~\ref{sec:matching}, we derive the spectral matching criterion for dark modes and apply it to illustrative examples in Sec.~\ref{sec:examples}. In Sec.~\ref{sec:entanglement}, we examine entanglement between synchronized qubits. Finally, in Sec.~\ref{sec:conc}, we summarize our results and discuss their implications.

\section{XX Model}
\label{sec:model}

We study an XX chain of $N$ qubits with open boundary conditions, site-dependent frequencies $\omega_i$, and nearest-neighbor exchange couplings $J_i$. Fixing $\hbar=1$, the Hamiltonian can be expressed as
\begin{equation}
H=-\frac12\sum_{i=1}^{N}\omega_i\,\sigma_z^{(i)}
+\frac12\sum_{i=1}^{N-1}J_i
\left(
\sigma_x^{(i)}\sigma_x^{(i+1)}
+\sigma_y^{(i)}\sigma_y^{(i+1)}
\right),
\label{eq:H}
\end{equation}
where $\sigma_x^{(i)}$, $\sigma_y^{(i)}$, and $\sigma_z^{(i)}$ are the Pauli operators acting on qubit $i$. We allow arbitrary real frequencies $\omega_i$ and coupling strengths $J_i$, and assume that every $J_i$ is nonzero so that the chain is connected. The uniform and resonant chain of Ref.~\cite{Cakmak2026} is the special case $\omega_i=\omega$ and $J_i=J$.

We use the basis states $|0\rangle=(1,0)^T$ and $|1\rangle=(0,1)^T$, representing the absence and presence of an excitation, respectively. With the convention $\sigma_\pm=(\sigma_x\mp i\sigma_y)/2$, the ladder operators become $\sigma_+^{(i)}=|1\rangle\langle0|$ and $\sigma_-^{(i)}=|0\rangle\langle1|$. The Hamiltonian can be written equivalently as
\begin{equation}
H=-\frac12\sum_{i=1}^{N}\omega_i\,\sigma_z^{(i)}
+\sum_{i=1}^{N-1}J_i
\left(
\sigma_+^{(i)}\sigma_-^{(i+1)}
+\sigma_-^{(i)}\sigma_+^{(i+1)}
\right).
\label{eq:Hladder}
\end{equation}
This form makes the action of the interaction explicit, i.e., an excitation is only exchanged between neighboring qubits without changing the total number of excitations. Accordingly, $[H,\hat N]=0$, where $\hat N=\sum_{i=1}^{N}\sigma_+^{(i)}\sigma_-^{(i)}$, and the Hamiltonian in Eq.~\eqref{eq:Hladder} decomposes into sectors with $k=0,1,\ldots,N$ excitations. The zero-excitation sector consists of a single state, the vacuum $|G\rangle=|0\rangle^{\otimes N}$, whose energy we set to zero by a constant shift of the Hamiltonian. In the one-excitation sector, spanned by the states $|j\rangle$ with $j$ being the label of the excited qubit, the Hamiltonian reads as follows
\begin{equation}
H^{(1)}=\sum_{j=1}^{N}\omega_j|j\rangle\langle j|
+\sum_{j=1}^{N-1}J_j
\left(
|j\rangle\langle j+1|
+|j+1\rangle\langle j|
\right).
\label{eq:H1}
\end{equation}
The one-excitation Hamiltonian takes the tight-binding form, with on-site energies $\omega_j$ and nearest-neighbor hopping amplitudes $J_j$. In the basis spanned by $\{|j\rangle\}$, it is represented by a real symmetric tridiagonal matrix.

For an eigenstate $|\psi\rangle=\sum_j c_j|j\rangle$ having energy $E$ relative to the zero-excitation state, the eigenvalue equation $H^{(1)}\vec c=E\vec c$ for $\Vec{c}=(c_1,\dots,c_N)^T$ can be expressed  component by component as
\begin{equation}
J_{j-1}c_{j-1}+\omega_jc_j+J_jc_{j+1}=Ec_j,
\quad J_0=J_N=0.
\label{eq:rec}
\end{equation}
This three-term recurrence relates each site's amplitude to those at the neighboring sites. Since every coupling is nonzero, specifying $E$ and $c_1$ fixes all subsequent amplitudes. Each eigenspace is therefore one dimensional, and the spectrum of $H^{(1)}$ is nondegenerate. The recurrence relation in Eq.~\eqref{eq:rec} also implies that an eigenvector cannot vanish at either endpoint or at two consecutive sites in the chain because such zeros would propagate through the chain and force the entire vector to vanish~\cite{Kay2010,Teschl2000}. Hence, any zero-amplitude sites of an eigenvector lie in the interior of the chain and are isolated. These properties of the eigenstates of $H^{(1)}$ will be useful when identifying the states protected from relaxation.

We now consider the case in which the XX chain is subjected to local amplitude-damping noise on a set of sites $M=\{m_1,\ldots,m_q\}$. Its density operator obeys the Lindblad master equation~\cite{Lindblad1976,Gorini1976}
\begin{equation}
\dot\rho=-i[H,\rho]
+\sum_{a=1}^{q}\gamma_a
\left(
L_a\rho L_a^\dagger
-\frac12\{L_a^\dagger L_a,\rho\}
\right),
\label{eq:lindblad}
\end{equation}
with the jump operators $L_a=\sigma_-^{(m_a)}$ and damping rates $\gamma_a>0$. Each jump removes an excitation from the corresponding qubit, taking it from $|1\rangle$ to $|0\rangle$. Even though coherent evolution preserves excitation number, the dissipation transfers population from a sector with $k$ excitations to one with $k-1$. The damping rates control the dissipative timescales and can affect the populations and coherences left after the transient. Their values, however, do not affect the subspace protected from the noise, which is determined by the Hamiltonian and the locations of the losses, as discussed in the following section.

\section{Decoherence-Free Subspace}
\label{sec:dfs}

A DFS supports unitary evolution despite the presence of dissipation~\cite{Zanardi1997,Lidar1998, Knill2000,Lidar2003}. First, the jump operators must act as scalars on this subspace. Due to the fact that each local lowering operator $\sigma_-^{(m)}$ has zero as its only eigenvalue, this requirement reduces here to the annihilation of every state in the DFS by every $L_a$. Second, the subspace must also remain invariant under the Hamiltonian $H$, so that coherent evolution cannot take a protected state outside it. Since $H$ is Hermitian, any such invariant subspace admits a basis of Hamiltonian eigenstates. Combining these observations, the DFS is spanned by the eigenstates of $H$ that are annihilated by every jump operator. In particular, it always contains the zero-excitation state $|G\rangle$. 

In the one-excitation sector, the annihilation condition has a simple interpretation in terms of eigenstate amplitudes. For an arbitrary state $|\psi\rangle=\sum_j c_j|j\rangle$, the operator $\sigma_-^{(m)}$ annihilates every basis state $|j\rangle$ with $j\ne m$, since qubit $m$ is then in $|0\rangle$, and maps $|m\rangle$ to $|G\rangle$. Therefore, $\sigma_-^{(m)}|\psi\rangle=c_m|G\rangle.$ A one-excitation eigenstate is protected from relaxation if and only if its amplitude vanishes at every damped site. We refer to these eigenstates as dark modes and denote them by $|D_\alpha\rangle$, with
\begin{equation}
H^{(1)}|D_\alpha\rangle=E_\alpha|D_\alpha\rangle,
\qquad
d_\alpha(m_a)=0
\label{eq:darkdef}
\end{equation}
where $m_a$ denotes the locations of the damped sites, $d_\alpha(j)=\langle j|D_\alpha\rangle$, and $\alpha=1,\ldots,r$. The integer $r$ is the dimension of the one-excitation dark subspace, or equivalently, the number of linearly independent dark modes in this sector. The nodal properties discussed in Sec.~\ref{sec:model} impose two restrictions on the loss locations. Damping either endpoint gives $r=0$, since no eigenvector of $H^{(1)}$ can vanish there. Likewise, damping two adjacent sites gives $r=0$, since an eigenvector cannot have two consecutive zero amplitudes. As a result, a nonzero $r$ implies a DFS containing at least one one-excitation dark mode alongside $|G\rangle$, and requires all damped sites to lie in the interior of the chain, with no two of them adjacent.

For excitation-preserving models whose Hamiltonians become noninteracting and number conserving under the Jordan-Wigner transformation, the multi-excitation eigenstates can be constructed from one-excitation eigenstates~\cite{Lieb1961,Takahashi1999}. A complete eigenbasis in each $k$-excitation sector can be generated from sets of $k$ orthonormal one-excitation eigenstates. The amplitudes in the ordered spin basis are Slater determinants, and the corresponding energies are sums of the constituent one-excitation energies. Since the XX chain considered in our work belongs to this class, its one-excitation dark modes provide the building blocks of the higher-excitation DFS sectors.

In particular, given two orthonormal one-excitation dark modes $|D_\alpha\rangle$ and $|D_\beta\rangle$, with $\alpha<\beta$, one can indeed construct the two-excitation dark state as
\begin{equation}
|D_{\alpha\beta}\rangle
=\sum_{1\le i<j\le N}
\Psi_{\alpha\beta}(i,j)\,|i,j\rangle,
\label{eq:slater}
\end{equation}
where $|i,j\rangle$ has excitations at sites $i$ and $j$, and
\begin{equation}
\begin{aligned}
\Psi_{\alpha\beta}(i,j)
=
\begin{vmatrix}
d_\alpha(i) & d_\beta(i)\\
d_\alpha(j) & d_\beta(j)
\end{vmatrix}
=d_\alpha(i)d_\beta(j)-d_\alpha(j)d_\beta(i).
\end{aligned}
\label{eq:slateramp}
\end{equation}
The energy of $|D_{\alpha\beta}\rangle$ is $E_{\alpha\beta}=E_\alpha+E_\beta$. If either occupied site is damped, both entries in the corresponding row are zero, so the determinant in Eq.~\eqref{eq:slateramp} vanishes. Hence, $|D_{\alpha\beta}\rangle$ has no component with an excited damped qubit and is annihilated by every jump operator. Higher-excitation dark states are constructed in the same way. For $k$ distinct one-excitation dark modes, the amplitude of each configuration of $k$-excited sites is a Slater determinant, with rows corresponding to the excited sites and columns to the chosen dark modes, and the energy of the resulting eigenstate is the sum of their one-excitation energies. If any excited site is damped, the corresponding row vanishes since every constituent dark mode has zero amplitude there, so such states are also dark. The dark states in the $k$-excitation sector is constructed by choosing $k$ of the $r$ dark modes, which yields $\binom{r}{k}$ independent states, and the dimension of the DFS is given as $\dim\mathcal H_{\mathrm{DFS}}=2^r$. The case that is of particular importance is $r=1$, for which $\mathcal H_{\mathrm{DFS}}=\operatorname{span}\{|G\rangle,|D_1\rangle\}$ has no multi-excitation component at all. As we discuss in Sec.~\ref{sec:sync}, this absence of higher-excitation dark states is  what allows the chain to exhibit generic synchronization.

In fact, the zero-excitation state $|G\rangle$, the dark modes $|D_\alpha\rangle$, and the higher-excitation dark states constructed from them span the DFS. In other words, any combination of bright modes, i.e., one-excitation eigenmodes with nonzero amplitude at one or more damped sites, cannot be used to construct a state in the DFS, not even through cancellations between several of their components.

\section{Long-Time Dynamics}
\label{sec:dynamics}

At long times, the decaying contributions to the density operator become negligible, and the surviving evolution is confined to the DFS. Since every jump operator annihilates this subspace and the Hamiltonian preserves it, the remaining dynamics is purely unitary,
\begin{equation}
\rho_{\mathrm{DFS}}(t)
=e^{-iHt}\rho^\infty e^{iHt}.
\label{eq:rhoDFS}
\end{equation}
Here, $\rho^\infty$ denotes the density operator of the whole chain within the DFS at the onset of the unitary regime, once the decaying contributions have become negligible. The time $t$ in Eq.~\eqref{eq:rhoDFS} is measured from this point. The populations and coherences of $\rho^\infty$ are determined by the initial state and the dissipative evolution during the transient, including any transfer of population and coherence from higher-excitation sectors into the DFS. 

Our main aim is to analyze the onset of stable quantum synchronization using the local expectation values of the qubits, $\langle\sigma_x(t)\rangle$. Since the observable $\sigma_x=\sigma_++\sigma_-$ can only change the excitation number by one, it probes coherences between neighboring excitation sectors. In particular, it is sensitive to the coherence between the zero-excitation dark state and a one-excitation dark mode, which is the only coherence within the DFS when $r=1$. This makes $\sigma_x$ a suitable local observable for identifying persistent oscillations and comparing their frequencies, amplitudes, and phases at different sites. To evaluate the expectation value, we follow the approach of Ref.~\cite{Cakmak2026} and work in an orthonormal basis $\{|\nu\rangle\}$ of the DFS consisting of eigenstates of $H$ with definite excitation number, $H|\nu\rangle=E_\nu|\nu\rangle$ and $\hat N|\nu\rangle=n_\nu|\nu\rangle$. Such a basis always exists because $H$ preserves the DFS and commutes with $\hat N$, and the dark states constructed in Sec.~\ref{sec:dfs} provide one explicitly. In this basis, Eq.~\eqref{eq:rhoDFS} reads
\begin{equation}
\langle\nu|\rho_{\mathrm{DFS}}(t)|\mu\rangle
=\rho^\infty_{\nu\mu}e^{-i(E_\nu-E_\mu)t},
\label{eq:DFScoherence}
\end{equation}
where $\rho^\infty_{\nu\mu}=\langle\nu|\rho^\infty|\mu\rangle$. As a consequence, the populations in this basis are constant, whereas the coherences acquire phases determined by the energy differences. Inserting this basis into $\operatorname{Tr}[\rho_{\mathrm{DFS}}(t)\sigma_x^{(\ell)}]$ gives
\begin{equation}
\begin{aligned}
\langle\sigma_x^{(\ell)}(t)\rangle_{\mathrm{DFS}}
={}&\sum_{\mu,\nu\in\mathrm{DFS}}
\rho^\infty_{\nu\mu}
\langle\mu|\sigma_x^{(\ell)}|\nu\rangle e^{-i(E_\nu-E_\mu)t}.
\end{aligned}
\label{eq:traceDFS}
\end{equation}
Since $\sigma_x^{(\ell)}=\sigma_+^{(\ell)}+\sigma_-^{(\ell)}$, its matrix elements vanish unless $n_\nu=n_\mu\pm1$.
Hermiticity implies that the contributions with $n_\nu=n_\mu-1$ are the complex conjugates of those with $n_\nu=n_\mu+1$. Combining them, we obtain
\begin{equation}
\langle\sigma_x^{(\ell)}(t)\rangle_{\mathrm{DFS}}
=2\operatorname{Re}
\sum_{\substack{\mu,\nu\in\mathrm{DFS}\\
n_\nu=n_\mu+1}}
\rho^\infty_{\nu\mu}X^{(\ell)}_{\mu\nu}
e^{-i(E_\nu-E_\mu)t},
\label{eq:formula}
\end{equation}
where $X^{(\ell)}_{\mu\nu}=\langle\mu|\sigma_-^{(\ell)}|\nu\rangle.$ Only the lowering part contributes to this matrix element because $|\nu\rangle$ contains one more excitation than $|\mu\rangle$. In fact, Eq.~\eqref{eq:formula} explicitly displays the factors that determine the asymptotic expectation values. For a nonzero energy difference, each contributing pair produces an oscillation at angular frequency $|E_\nu-E_\mu|$, with amplitude $2|\rho^\infty_{\nu\mu}X^{(\ell)}_{\mu\nu}|$.

It is worth noting that the same derivation applies to $\sigma_y$, whose expectation value is obtained by replacing the real part in Eq.~\eqref{eq:formula} with the imaginary part. Thus, any fixed qubit observable in the $xy$ plane probes the same neighboring-sector coherences and allowed transition frequencies. By contrast, $\sigma_z$ preserves the excitation number and probes populations and coherences within the same excitation sector. In particular, if $r=1$, its asymptotic expectation value becomes time independent, even when the expectation values of $\sigma_x$ and $\sigma_y$ oscillate persistently. Other observables can be analyzed within the same DFS framework, but their synchronization conditions depend on the  coherences they probe.

As the dark-state energies are additive, distinct transitions can share the same energy difference. For example, given $|D_\alpha\rangle$ and $|D_\beta\rangle$ with $\alpha<\beta$, the transitions connecting $|G\rangle$ to $|D_\alpha\rangle$ and $|D_\beta\rangle$ to
$|D_{\alpha\beta}\rangle$ satisfy
\begin{equation}
E_{D_\alpha}-E_G
=E_{D_{\alpha\beta}}-E_{D_\beta}
=E_\alpha,
\label{eq:sharedfrequency}
\end{equation}
and correspond to the same angular frequency $|E_\alpha|$ when $E_\alpha\ne0$. Indeed, the same energy difference arises whenever mode $\alpha$ is added to any set of occupied dark modes that does not already contain it. Two features of Eq.~\eqref{eq:formula} follow from this particular point. First, several distinct coherences can contribute at the same frequency, so their contributions must be added to obtain the expectation value of a qubit in the chain. Second, the relative signs of these contributions enter through the matrix elements $X^{(\ell)}_{\mu\nu}$,  and thus their total amplitudes can vary from qubit to qubit. This means that identical transition frequencies alone cannot ensure that the qubit expectation values remain proportional, a point that proves essential in the synchronization analysis of Sec.~\ref{sec:sync}.

In general, the density operator $\rho^\infty$ depends on the dissipative transient, during which population and coherence can be transferred from higher-excitation sectors into the DFS. For initial states with at most one excitation, however, this transfer can only add to the vacuum population, allowing the asymptotic state to be determined explicitly. Let us consider a pure initial state
\begin{equation}
|\psi_0\rangle=c_G|G\rangle+\sum_{j=1}^{N}c_j|j\rangle,
\label{eq:psi0}
\end{equation}
with $|c_G|^2+\sum_j|c_j|^2=1$. Expanding the one-excitation part in the eigenmodes of $H^{(1)}$, we can separate its dark and bright components as
\begin{equation}
\sum_j c_j|j\rangle
=\sum_{\alpha=1}^{r}a_\alpha|D_\alpha\rangle+|B\rangle,
\label{eq:split}
\end{equation}
where $a_\alpha=\langle D_\alpha|\psi_0\rangle =\sum_{j=1}^{N}d_\alpha(j)c_j$ is the initial amplitude of dark mode $\alpha$, and $|B\rangle$ collects the contributions of the remaining bright modes, which have nonvanishing amplitude at one or more damped sites. Since each dark mode is an eigenstate of $H^{(1)}$ annihilated by every jump operator, its component retains its initial magnitude and acquires a phase determined by its energy. The state of the chain never contains more than one excitation, so removing that excitation leaves the chain in the vacuum state $|G\rangle$. The bright component eventually decays completely, while the coherences among $|G\rangle$ and the dark modes hold their magnitudes and acquire phases set by the corresponding energy differences. To retain the time convention of Eq.~\eqref{eq:rhoDFS}, let $t_0$ be the elapsed time from the beginning of the dissipative dynamics to the onset of the unitary regime. The surviving coherent component is 
\begin{equation}
|\psi_D(t)\rangle
=c_G|G\rangle
+\sum_{\alpha=1}^{r}
a_\alpha e^{-iE_\alpha(t_0+t)}|D_\alpha\rangle.
\label{eq:psiD}
\end{equation}
Its squared norm is
$|c_G|^2+\sum_\alpha|a_\alpha|^2=1-p_{\mathrm{loss}}$,
where $p_{\mathrm{loss}}
=\langle B|B\rangle
=1-|c_G|^2-\sum_{\alpha=1}^{r}|a_\alpha|^2.$
The asymptotic density operator is
\begin{equation}
\rho_{\mathrm{DFS}}(t)
=|\psi_D(t)\rangle\langle\psi_D(t)|
+p_{\mathrm{loss}}|G\rangle\langle G|.
\label{eq:rhoDFSexact}
\end{equation}
The second term restores the population transferred to the vacuum during relaxation and ensures normalization. Reading off the matrix elements at $t=0$ gives
\begin{equation}
\begin{aligned}
\rho^\infty_{D_\alpha G}
&=a_\alpha c_G^*e^{-iE_\alpha t_0},
\qquad
\rho^\infty_{GG}
=1-\sum_{\alpha=1}^{r}|a_\alpha|^2,\\
\rho^\infty_{D_\alpha D_\beta}
&=a_\alpha a_\beta^*
e^{-i(E_\alpha-E_\beta)t_0}.
\end{aligned}
\label{eq:rhoinfexact}
\end{equation}
Thus, the asymptotic state is fixed by the initial state and the structure of the dark modes, without having to solve the dissipative transient dynamics. The local expectation values can be obtained directly. Within the accessible DFS sectors, $\sigma_x^{(\ell)}$ connects only the vacuum and the one-excitation dark modes, with $\langle G|\sigma_-^{(\ell)}|D_\alpha\rangle=d_\alpha(\ell)$. Equation~\eqref{eq:formula} reduces to
\begin{equation}
\langle\sigma_x^{(\ell)}(t)\rangle_{\mathrm{DFS}}
=2\operatorname{Re}\biggl[
c_G^*\sum_{\alpha=1}^{r}a_\alpha d_\alpha(\ell)
e^{-iE_\alpha(t_0+t)}
\biggr].
\label{eq:expectation01}
\end{equation}
Therefore, the asymptotic expectation values are
independent of the positive damping rates and are determined entirely by the initial coefficients and the dark-mode amplitudes and energies. For initial states containing two or more excitations, relaxation can also feed excited states within the DFS. Then, the surviving populations and coherences can depend on the transient and cannot be inferred from the initial dark-mode overlaps.

\section{Stable Synchronization}
\label{sec:sync}

Having characterized the asymptotic dynamics, we can now specify what we mean by synchronization in this setting. We say that two qubits $\ell$ and $\ell'$ exhibit stable synchronization if their asymptotic expectation values are nonconstant and remain proportional at all times,
\begin{equation}
\langle\sigma_x^{(\ell)}(t)\rangle_{\mathrm{DFS}}
=C_{\ell\ell'}\,
\langle\sigma_x^{(\ell')}(t)\rangle_{\mathrm{DFS}},
\label{eq:syncdef}
\end{equation}
with a real locking constant $C_{\ell\ell'}\neq0$. A positive constant describes in-phase oscillations and a negative one describes antiphase oscillations, while $\lvert C_{\ell\ell'} \rvert$ fixes their amplitude ratio. The locking is stable in the sense that this proportionality persists indefinitely during the asymptotic unitary evolution. In general, the relation given in Eq.~\eqref{eq:syncdef} holds only for some initial states. For this reason, we call a pair of qubits generically synchronized if Eq.~\eqref{eq:syncdef} holds with the same constant $C_{\ell\ell'}$ for every initial state producing persistent oscillations at both of the two chosen sites. This stronger requirement guarantees that the locking does not rely on a particular choice of the initial state. Hence, generic synchronization is enforced by the structure of the DFS, i.e., by the frequencies, couplings, and damping locations that define the chain.

To determine what is required for generic synchronization, let us now return to Eq.~\eqref{eq:formula} and recall that the only site dependent contribution is coming from the matrix element $X^{(\ell)}_{\mu\nu}$. Therefore, generic synchronization requires
\begin{equation}
X^{(\ell)}_{\mu\nu}
=C_{\ell\ell'}X^{(\ell')}_{\mu\nu},
\label{eq:termwise}
\end{equation}
for every pair with $n_\nu=n_\mu+1$, with the same constant $C_{\ell\ell'}$ for all pairs. This includes equal energy pairs, since mixing their stationary preparation with an oscillating one requires the constant contribution to obey the same proportionality. Conversely, Eq.~\eqref{eq:termwise} makes the expectation values proportional term by term for every initial state. Thus, for qubits capable of persistent oscillations, generic synchronization is equivalent to the proportionality of their local transition amplitudes in the DFS.

We now obtain a necessary and sufficient condition for generic synchronization of the edge qubits, which requires neither uniform frequencies and couplings nor spatial symmetry of the chain. For $r=0$, the DFS contains only $|G\rangle$, so all asymptotic expectation values of $\sigma_x$ vanish and stable synchronization is absent. For $r=1$, the DFS is covered by $\operatorname{span}\{|G\rangle,|D_1\rangle\}$, which forms a single pair of dark eigenstates whose excitation numbers differ by one. The transition amplitudes of this pair at the two endpoints are $X^{(1)}_{GD_1}=d_1(1)$ and $X^{(N)}_{GD_1}=d_1(N)$, and since neither endpoint amplitude can vanish, Eq.~\eqref{eq:termwise} is satisfied with the nonzero constant 
\begin{equation}
C_{1N}=d_1(1)/d_1(N),
\end{equation}
independent of the initial state. Provided $E_1\neq0$, any state with $\rho^\infty_{D_1G}\neq0$ produces persistent oscillations at both endpoints. A single dark mode of nonzero energy is sufficient for generic edge synchronization.

To establish that the above condition is also necessary, say $r\geq2$ and choose two dark modes $|D_\alpha\rangle$ and $|D_\beta\rangle$, with $\alpha<\beta$, with their two-excitation dark state $|D_{\alpha\beta}\rangle$. The transitions connecting the dark states $|G\rangle$ to $|D_\alpha\rangle$ and $|D_\beta\rangle$ to $|D_{\alpha\beta}\rangle$ have the same energy difference, $E_\alpha$, and satisfy Eq.~\eqref{eq:termwise} with the same constant. While the former transition yields
\begin{equation}
X^{(1)}_{GD_\alpha}=d_\alpha(1),
\qquad
X^{(N)}_{GD_\alpha}=d_\alpha(N),
\label{eq:edge_single}
\end{equation}
evaluating the latter transition result in
\begin{equation}
\begin{aligned}
X^{(1)}_{D_\beta D_{\alpha\beta}}
&=\sum_{j=2}^{N}d_\beta(j)\Psi_{\alpha\beta}(1,j)=d_\alpha(1),\\
X^{(N)}_{D_\beta D_{\alpha\beta}}
&=\sum_{j=1}^{N-1}d_\beta(j)\Psi_{\alpha\beta}(j,N)=-d_\alpha(N).
\end{aligned}
\label{eq:edge_double}
\end{equation}
The minus sign at the right endpoint is a direct consequence of the anti-symmetrization coefficients and does not stem from the spatial symmetry of the chain. Enforcing Eq.~\eqref{eq:termwise} to both transitions would require
\begin{equation}
d_\alpha(1)=C_{1N}d_\alpha(N),
\qquad
d_\alpha(1)=-C_{1N}d_\alpha(N)
\label{eq:edge_contradiction}
\end{equation}
at once, which is clearly impossible for any nonzero $d_\alpha(1)$ and $d_\alpha(N)$. Hence, $r\geq2$ excludes generic edge synchronization, although synchronization can still occur for particular initial states. As a result, we conclude that the edge qubits exhibit generic stable synchronization if and only if $r=1$ and $E_1\neq0$.

The sufficiency part of the proof is in fact not restricted to the edges of the chain. For $r=1$, Eq.~\eqref{eq:formula} gives
\begin{equation}
\langle\sigma_x^{(\ell)}(t)\rangle_{\mathrm{DFS}}
=2d_1(\ell)\operatorname{Re}\left[\rho^\infty_{D_1G}e^{-iE_1t}\right],
\label{eq:sync_r1}
\end{equation}
so that every site with $d_1(\ell)\neq0$ has the same time dependence, and any two of them are generically synchronized with the constant ratio
\begin{equation}
C_{\ell\ell'}=d_1(\ell)/d_1(\ell') ,
\label{eq:sync_ratio}
\end{equation}
independently of the initial state and of the dissipative transient dynamics. The relative signs of the dark-mode amplitudes decide whether the oscillations are in phase or in antiphase, and their magnitudes set the amplitude. A unique dark mode of nonzero energy therefore synchronizes the entire support of that mode, and not merely the two ends of the chain.

\section{Dark Modes of the Inhomogeneous XX Chain}
\label{sec:matching}

Sections~\ref{sec:dfs} to \ref{sec:sync} reduced generic edge synchronization to the condition $r=1$ with $E_1\neq0$, without using explicit eigenmodes. It remains to determine which frequencies and couplings produce this structure, that is, how many one-excitation eigenmodes vanish at every damped site.

For the uniform and resonant XX chain, this question has a purely arithmetic answer. The one-excitation modes labeled by $n=1,\ldots,N$ has amplitudes proportional to $\sin[nj\pi/(N+1)]$ at site $j$. It vanishes at every damped site precisely when $nm_a/(N+1)\in\mathbb{Z}$ for all $a=1,\ldots,q$. Counting the mode indices that satisfy these conditions yields $r=\gcd(m_1,\ldots,m_q,N+1)-1$, as demonstrated in Ref.~\cite{Cakmak2026}. Thus, $r=1$ is equivalent to $\gcd(m_1,\ldots,m_q,N+1)=2$. Note that the count is solely fixed by the chain length and the damping locations and completely independent of the Hamiltonian parameters. Moreover, the inversion symmetry of the uniform chain fixes the endpoint amplitude ratio of each mode to $\pm1$. Provided a unique dark mode of nonzero energy survives, the edge qubits oscillate with equal amplitudes, either in phase or in antiphase, if $\rho^\infty_{D_1G}\neq0$.

For an inhomogeneous chain, the dark mode count is no longer determined by the chain length and damping locations alone. The eigenmodes do not have a simple sinusoidal form, and their nodes depend on the site frequencies and exchange couplings. However, the darkness condition remains the same, since the amplitudes must vanish at every damped site, but it must now be imposed on the eigenvectors that are not known beforehand. 
In the following, we determine the dark modes by imposing the conditions for a dark mode in the eigenvalue equation for the one-excitation sector Hamiltonian. A damped site separates the equations on its two sides into eigenvalue problems for the two segments, while the condition at the damped site yields a single relation between the amplitudes on its two neighboring sites~\cite{Michor2004,Parlett1998,Parlett2016}.

\subsection{Single-Site Noise}

Let us first consider a single damped site $m$ in the chain, with $1<m<N$, which divides the chain into a left segment $L=\{1,\ldots,m-1\}$ and a right segment $R=\{m+1,\ldots,N\}$. Let $h_L$ and $h_R$ be the corresponding one-excitation Hamiltonians, obtained from Eq.~\eqref{eq:H1} by restricting it to the sites of each segment, so that $h_L$ contains the frequencies $\omega_1,\ldots,\omega_{m-1}$ and the couplings $J_1,\ldots,J_{m-2}$, while $h_R$ contains $\omega_{m+1},\ldots,\omega_N$ and $J_{m+1},\ldots,J_{N-1}$. The frequency $\omega_m$ of the damped qubit and the two couplings $J_{m-1}$ and $J_m$ that join it to its neighbors belong to neither segment. In this decomposition, the one-excitation Hamiltonian has the block form
\begin{equation}
H^{(1)}=
\begin{pmatrix}
h_L & J_{m-1}\hat r_L & 0\\
J_{m-1}\hat r_L^{T} & \omega_m & J_m\hat\ell_R^{T}\\
0 & J_m\hat\ell_R & h_R
\end{pmatrix}.
\label{eq:H1_blocks}
\end{equation}
For an eigenvector written as $\vec c=(\vec c_L,c_m,\vec c_R)$, the eigenvalue equation $H^{(1)}\vec c=E\vec c$ splits into
\begin{equation}
\begin{aligned}
h_L\vec c_L+J_{m-1}c_m\,\hat r_L&=E\vec c_L,\\
J_{m-1}c_{m-1}+\omega_mc_m+J_mc_{m+1}&=Ec_m,\\
h_R\vec c_R+J_mc_m\,\hat\ell_R&=E\vec c_R,
\end{aligned}
\label{eq:blocks}
\end{equation}
where $\hat r_L$ and $\hat\ell_R$ are the unit vectors selecting the last site of $L$ and the first site of $R$, respectively. Imposing the darkness condition $c_m=0$ removes the coupling terms from the first and third lines of Eq.~\eqref{eq:blocks}, which become independent eigenvalue problems for the two segments,
\begin{equation}
h_L\vec c_L=E\vec c_L,
\qquad
h_R\vec c_R=E\vec c_R.
\label{eq:segments}
\end{equation}
The second line loses its diagonal term and reduces to a single scalar relation between the amplitudes on the two sites adjacent to the damped one, that is,
\begin{equation}
J_{m-1}c_{m-1}+J_mc_{m+1}=0.
\label{eq:scalar}
\end{equation}
It is important to note that neither $\vec c_L$ nor $\vec c_R$ can vanish identically. If $\vec c_L$ were zero, its last entry $c_{m-1}$ would vanish together with $c_m$, giving two consecutive zero amplitudes and forcing the entire eigenvector to vanish, and the same argument applies to $\vec c_R$ as well. Then, Eq.~\eqref{eq:segments} dictates that both $\Vec{c}_L$ and $\Vec{c}_R$ are a nonzero eigenvectors of its own segment at the same energy $E$. Thus, a necessary condition for a dark mode of energy $E$ is that $E$ belong to the spectra of both $h_L$ and $h_R$.

In fact, this spectral matching condition is also a sufficient one. Both the left and the right segment Hamiltonians are real symmetric tridiagonal matrices with nonzero nearest-neighbor couplings, so the properties of the matrices discussed in Sec.~\ref{sec:model} apply to both segments separately. Their spectra are nondegenerate, and their eigenvectors cannot vanish at either endpoint. Now, let $\vec u$ and $\vec v$ be the normalized eigenvectors of $h_L$ and $h_R$ that have $E$ as their common eigenvalue. Their amplitudes $u_{m-1}$ and $v_{m+1}$ at the sites adjacent to the damped qubit are both nonzero. A full-chain vector consisting of arbitrary superpositions of these two eigenvectors can be written in the following form $\vec c=(A\vec u,0,B\vec v)$. Naturally, such a state must satisfy Eq.~\eqref{eq:segments} and the remaining equation involving the sites around the damped site, Eq.~\eqref{eq:scalar}, which imposes the condition
\begin{equation}
\frac{B}{A}
=-\frac{J_{m-1}u_{m-1}}{J_m v_{m+1}}.
\label{eq:relative_weights}
\end{equation}
Note that the right-hand side of the above equation is always nonzero and finite, and fixes the relative amplitude of the two segments. After normalization, the resulting vector is an eigenstate of $H^{(1)}$ with energy $E$ and is a dark mode. In other words, a dark mode with energy $E$ exists if and only if left and right segment Hamiltonians share $E$ as a common eigenvalue. Moreover, the nondegeneracy of the segment spectra fixes $\vec u$ and $\vec v$ up to an overall phase, and Eq.~\eqref{eq:relative_weights} fixes their relative weight. Each common eigenvalue produces exactly one linearly independent dark mode. As a consequence, the number of dark modes is given by
\begin{equation}
r=\left|
\operatorname{spec}(h_L)\cap
\operatorname{spec}(h_R)
\right|,
\label{eq:dark_count_single}
\end{equation}
where the right-hand side denotes the number of eigenvalues shared by the two segments.

The dark-mode counting rule above also clarifies the roles of the different parameters of the chain. The frequency $\omega_m$ of the damped qubit appears neither in a segment Hamiltonian nor in Eq.~\eqref{eq:scalar}, so changing it leaves the number, energies, and spatial profiles of the dark modes unchanged. The neighboring couplings to the damped site, $J_{m-1}$ and $J_m$, also lie outside the segment Hamiltonians, therefore varying them do not change the common eigenvalues, and thus the number of dark modes. They only enter in the discussion through Eq.~\eqref{eq:relative_weights}, which fixes the relative amplitudes of the two segments, i.e. the spatial profile of the dark mode. Overall, while the segment spectra determine whether generic edge synchronization is supported, the couplings right next to the damped site allow the spatial weights of the dark modes on the two sides to be adjusted.

An interesting consequence of the latter is that it allows us to control the amplitude of the synchronized oscillations for $r=1$. In Sec.~\ref{sec:sync}, we already showed that for a unique dark mode of nonzero energy all sites on its support are synchronized. Assuming the unique dark mode is $\vec d_1=(A\vec u,0,B\vec v)$, two sites within the same segment have a locking constant determined solely by this segment's eigenvector amplitudes, since the common factor $A$ or $B$ cancel. Yet, for the endpoints,
Eq.~\eqref{eq:relative_weights} gives
\begin{equation}
C_{1N}
=\frac{A u_1}{B v_N}
=-\frac{J_m}{J_{m-1}}
\frac{u_1v_{m+1}}{u_{m-1}v_N}.
\label{eq:edge_ratio_segments}
\end{equation}
For positive $J_{m-1}$ and $J_m$, the oscillations are in phase when $u_1/u_{m-1}$ and $v_{m+1}/v_N$ have opposite signs, and in antiphase when these ratios have the same sign. With the segment Hamiltonians held fixed, varying the magnitude of $J_m/J_{m-1}$ tunes the magnitude of $C_{1N}$ without changing the spectral matching condition or the dark-mode energy. The inhomogeneous chain allows synchronized edge oscillations with an adjustable amplitude ratio, whereas inversion symmetry fixes it to $\pm1$ in the uniform chain.

The spectral matching condition in Eq.~(\ref{eq:dark_count_single}) can also be expressed algebraically through the characteristic polynomials of the two segments,
\begin{equation}
P_L(E)=\det(EI-h_L),
\quad
P_R(E)=\det(EI-h_R).
\label{eq:segment_polynomials}
\end{equation}
Their common roots correspond precisely to the dark-mode energies. Let $G(E)=\gcd(P_L,P_R)$ denote their greatest common polynomial factor. Since both segment Hamiltonians have nondegenerate spectra, each common root appears exactly once, giving
\begin{equation}
G(E)=\prod_{\alpha=1}^{r}(E-E_\alpha),
\quad
r=\deg G,
\label{eq:polynomial_gcd}
\end{equation}
where $E_\alpha$ are the dark-mode energies. Thus, the degree of $G(E)$ determines the number of dark modes, while its roots give their energies. The polynomial greatest common divisor can be evaluated using the Euclidean algorithm, providing an algebraic means of determining the dark-mode count and energies without explicitly diagonalizing the segment Hamiltonians.

\subsection{Multi-Site Noise}

Let us next consider several damped sites in the chain, $M=\{m_1,\ldots,m_q\}$, ordered as $m_1<\cdots<m_q$. We assume that all damped sites are interior and that no two are adjacent, since otherwise $r=0$, as discussed in Sec.~\ref{sec:dfs}. In this case, the damped sites divide the chain into $q+1$ segments,
\begin{equation}
S_s=\{m_s+1,\ldots,m_{s+1}-1\},
\quad s=0,\ldots,q,
\label{eq:segments_multiple}
\end{equation}
with corresponding one-excitation Hamiltonians $h_s$, and $m_0=0$ and $m_{q+1}=N+1$. Forcing the amplitudes to vanish at every damped site reduces the eigenvalue equation on each segment to $h_s\vec c_s=E\vec c_s$. Similar to the single-site noise case, for a dark mode to exist there must be at least one eigenvalue that belong to the spectra of all $q+1$ segment Hamiltonians. Conversely, suppose that $E$ is an eigenvalue of every $h_s$, and let $\vec u_s$ be the corresponding normalized eigenvector of $h_s$. Then, it is again possible to construct a full-chain vector by superposing each segment eigenvector with certain weights, $\vec c_s=A_s\vec u_s$, and $c_{m_a}=0$ at every damped site. Generalization of Eq.~(\ref{eq:relative_weights}) to the present case, involves the two segments adjacent to the damped site $m_a$, $S_{a-1}$ on its left and $S_a$ on its right, and fixes their relative amplitudes,
\begin{equation}
\frac{A_a}{A_{a-1}}
=-\frac{J_{m_a-1}u_{a-1}(m_a-1)}
{J_{m_a}u_a(m_a+1)},
\quad a=1,\ldots,q.
\label{eq:multiple_matching}
\end{equation}
Here, $u_{a-1}(m_a-1)$ is the amplitude at the last site of the left segment $S_{a-1}$, and $u_a(m_a+1)$ is the amplitude at the first site of the right segment $S_a$. For any $A_0\neq0$, Eq.~\eqref{eq:multiple_matching} fixes the remaining amplitudes successively, and the resulting vector is a dark mode of the full chain.

Thus, an energy supports a dark mode if and only if it is shared by every segment spectrum. The
nondegeneracy of the segment spectra and the fixed relative amplitudes in Eq.~\eqref{eq:multiple_matching} ensure that each common eigenvalue produces exactly one  dark mode. Consequently,
\begin{equation}
r=\left|
\bigcap\nolimits_{s=0}^{q}\operatorname{spec}(h_s)
\right|.
\label{eq:dark_count_multiple}
\end{equation}

The characteristic-polynomial formulation extends to the multi-site noise case. Defining $P_s(E)=\det(EI-h_s)$ for each segment, the dark-mode counting rule in Eq.~(\ref{eq:dark_count_multiple}) can be expressed as
\begin{equation}
G(E)=\gcd(P_0,\ldots,P_q),
\qquad r=\deg G.
\label{eq:polynomial_gcd_multiple}
\end{equation}
Consequently, generic stable synchronization of the edge qubits occurs if and only if $G(E)$ is a first-degree polynomial with a nonzero root.

For a fixed Hamiltonian, adding damped sites imposes further zero-amplitude conditions on the one-excitation eigenmodes and, as a result, cannot increase $r$. Moreover, a segment consisting of a single site restricts the chain to at most one dark mode. Such a mode exists precisely when the frequency of that site is an eigenvalue of every other segment Hamiltonian.

\subsection{Recovering the Uniform-Chain Result}
\label{sec:uniform_recovery}

We now show how the polynomial counting rule recovers the integer gcd criterion of Ref.~\cite{Cakmak2026} for the uniform and resonant chain. For $\omega_i=\omega$ and $J_i=J\neq0$, the segments differ only in their lengths $N_s=|S_s|$. A segment of length $N_s$ has eigenvalues
\begin{equation}
E_{s,p}=\omega+2J\cos\!\left(\frac{p\pi}{N_s+1}\right),
\quad p=1,\ldots,N_s.
\label{eq:uniform_levels}
\end{equation}
Since the cosine decreases strictly between zero and $\pi$, equal energies necessitate equal angles. Finding a shared eigenvalue amounts to finding a fraction $p/(N_s+1)$ that occurs in every segment. Let $g=\gcd(N_0+1,\ldots,N_q+1)$ be the largest positive integer that divides every $N_s+1$. Every segment then contains the fractions $n/g$, where $n=1,\ldots,g-1$. Indeed, the fraction $n/g$ occurs in segment $s$ at the mode index $p=n(N_s+1)/g$, which is an integer between $1$ and $N_s$. Conversely, any fraction common to all segments can be written in lowest terms with a denominator that divides every $N_s+1$, and hence divides $g$, so it is already one of the fractions listed above. The segments share exactly $g-1$ eigenvalues, each appearing once in the polynomial gcd,
\begin{equation}
G(E)=\prod_{n=1}^{g-1}
\left[E-\omega-2J\cos\!\left(\frac{n\pi}{g}\right)\right],
\label{eq:uniform_polynomial_gcd}
\end{equation}
so that $r=g-1$. For $g=1$, the product is empty, i.e., $G(E)=1$, and $r=0$. This count can also be expressed in terms of the damping locations. The integers $N_s+1=m_{s+1}-m_s$ are the gaps between successive positions, with $m_0=0$ and $m_{q+1}=N+1$. Adding these gaps gives the positions $m_1,\ldots,m_q,N+1$. Any integer dividing all the gaps also divides all these positions. Conversely, an integer dividing all the positions also divides their differences. As a consequence, the gaps and the positions have the same greatest common divisor, and we have
\begin{equation}
r=\gcd(m_1,\ldots,m_q,N+1)-1.
\label{eq:uniform_gcd}
\end{equation}
This recovers the counting rule derived in Ref.~\cite{Cakmak2026} from the full-chain eigenmodes. When $g=2$, the only common fraction is $n/g=1/2$. The unique dark-mode energy  is $E_1=\omega+2J\cos(\pi/2)=\omega$, so $G(E)=E-\omega$. As a result, generic edge synchronization occurs if and only if $\gcd(m_1,\ldots,m_q,N+1)=2$ and $\omega\neq0$. The segments of a uniform chain share an energy if and only if their lengths plus one have a nontrivial common divisor.

\section{Illustrative Examples}
\label{sec:examples}

We now apply the spectral matching condition to selected chains, for which the conditions for dark modes and the resulting synchronized dynamics can be obtained explicitly. We begin with the simplest chain that supports a nontrivial DFS.

\subsection{Three-Qubit Chain}

The smallest chain that supports a one-excitation dark mode in the presence of local relaxation has $N=3$, with damping on the middle qubit, $M=\{2\}$. The two segments are the single sites $L=\{1\}$ and $R=\{3\}$, with 
\begin{equation}
\operatorname{spec}(h_L)=\{\omega_1\},
\qquad
\operatorname{spec}(h_R)=\{\omega_3\}.
\label{eq:N3_spectra}
\end{equation}
A shared eigenvalue exists if and only if the two edge qubits in the chain are resonant, that is,
\begin{equation}
\omega_1=\omega_3= E_1.
\label{eq:N3_condition}
\end{equation}
Since each segment has only one eigenvalue, this resonance condition yields $r=1$, whereas otherwise $r=0$. The scalar relation in Eq.~(\ref{eq:scalar}), $J_1d_1(1)+J_2d_1(3)=0$, together with normalization, determines the unique one-excitation dark mode completely,
\begin{equation}
|D_1\rangle
=\frac{J_2|1\rangle-J_1|3\rangle}
{\sqrt{J_1^2+J_2^2}}.
\label{eq:N3_mode}
\end{equation}
The edge qubits of the chain are generically synchronized, with locking constant
\begin{equation}
C_{13}=-J_2/J_1.
\label{eq:N3_ratio}
\end{equation}

Elementary as it is, this example already clearly separates the roles of the chain parameters. Resonance between the edge qubits decides whether a dark mode exists at all, whereas the frequency $\omega_2$ of the damped qubit enters neither Eq.~\eqref{eq:N3_condition} nor Eq.~\eqref{eq:N3_mode} and may be detuned arbitrarily without affecting the dark mode or its energy. This frequency influences the bright modes and the approach to the asymptotic regime. The couplings $J_1$ and $J_2$ in turn fix the spatial profile of the dark mode and the locking constant, whose magnitude can take any nonzero value, with antiphase oscillations for positive couplings. The same damping configuration in the uniform chain yields $r=\gcd(2,4)-1=1$ as well, but there equal couplings fix $C_{13}=-1$. We can see that inhomogeneity in this case does not create the synchronization. Rather, it makes the ratio of the oscillation amplitudes adjustable.

In particular, for the initial state $|\psi_0\rangle=|+\rangle_1|0\rangle_2|0\rangle_3$ with $|+\rangle=(|0\rangle+|1\rangle)/\sqrt{2}$, the vacuum amplitude and the overlap with the mode $\ket{D_1}$ are given by $c_G=1/\sqrt{2}$ and $a_1=J_2/\sqrt{2(J_1^2+J_2^2)}$, respectively. Equation~\eqref{eq:expectation01} gives
\begin{equation}
\begin{aligned}
\langle\sigma_x^{(1)}(t)\rangle_{\mathrm{DFS}}
&=\frac{J_2^2}{J_1^2+J_2^2}
\cos\!\big[E_1(t_0+t)\big],\\
\langle\sigma_x^{(3)}(t)\rangle_{\mathrm{DFS}}
&=-\frac{J_1J_2}{J_1^2+J_2^2}
\cos\!\big[E_1(t_0+t)\big],
\end{aligned}
\label{eq:N3_expectations}
\end{equation}
while $\langle\sigma_x^{(2)}(t)\rangle_{\mathrm{DFS}}=0$, with $t_0+t$ the elapsed time since the initial preparation. For positive coupling strengths $J_1$ and $J_2$, the edge qubits in the chain oscillate in antiphase at the angular frequency $|E_1|$ with the amplitude ratio $J_2/J_1$, in agreement with Eq.~\eqref{eq:N3_ratio}, and this asymptotic evolution follows from the initial state and the dark mode alone, without solving the dissipative transient.

\subsection{Four-Qubit Chain}

For a uniform chain, generic edge synchronization requires $\gcd(m_1,\ldots,m_q,N+1)=2$, which cannot possibly hold for even $N$, since $N+1$ is then odd. Therefore, no uniform chain of even length supports generic edge synchronization, irrespective of the damping configuration. Nonetheless, frequency tuning can overcome this obstruction. The simplest chain of even length in which this is possible has $N=4$ with damping on the second qubit, $M=\{2\}$. The undamped segments here are the single site $L=\{1\}$ and the dimer $R=\{3,4\}$, with spectra
\begin{equation}
\operatorname{spec}(h_L)=\{\omega_1\},
\qquad
\operatorname{spec}(h_R)=\{E_-,E_+\},
\label{eq:N4_spectra}
\end{equation}
where the eigenvalues of the right segment read
\begin{equation}
E_\pm=\frac{\omega_3+\omega_4}{2}
\pm\sqrt{\left(\frac{\omega_3-\omega_4}{2}\right)^2+J_3^2}.
\label{eq:N4_dimer_levels}
\end{equation}
Since the left segment Hamiltonian possesses only a single eigenvalue, at most one eigenvalue can be shared, so $r\leq 1$ for any choice of parameters. The characteristic polynomials of the two segments are $P_L(E)=E-\omega_1$ and $P_R(E)=(E-\omega_3)(E-\omega_4)-J_3^2$. Therefore, a common root exists if and only if $P_R(\omega_1)=0$, yielding
\begin{equation}
(\omega_1-\omega_3)(\omega_1-\omega_4)=J_3^2.
\label{eq:N4_condition}
\end{equation}
When this condition holds, we have $G(E)=E-\omega_1$ and the unique dark mode energy is $E_1=\omega_1$.
Thus, tuning the frequency of the first qubit to either of the dimer energies is sufficient to produce a unique dark mode. On the other hand, if all undamped qubits are resonant, the left-hand side of Eq.~\eqref{eq:N4_condition} vanishes. Since $J_3\neq0$ by assumption, changing the exchange couplings alone cannot produce a dark mode in that case. The dark mode follows directly from the segment eigenvector and the scalar relation at the damped site.
Within the dimer,
$J_3d_1(4)=(E_1-\omega_3)d_1(3)$, while the scalar relation reads $J_1d_1(1)+J_2d_1(3)=0$. Hence,
\begin{equation}
|D_1\rangle\propto
J_2|1\rangle-J_1|3\rangle
-\frac{J_1(E_1-\omega_3)}{J_3}|4\rangle.
\label{eq:N4_mode_general}
\end{equation}
The matching condition ensures that $E_1-\omega_3$ is nonzero, so this dark mode has nonvanishing amplitude at every undamped site. These three qubits are  generically synchronized, with locking constants
\begin{equation}
\begin{aligned}
C_{13}&=-\frac{J_2}{J_1},
\qquad
C_{34}=\frac{J_3}{E_1-\omega_3},\\
C_{14}&=C_{13}C_{34}
=-\frac{J_2J_3}{J_1(E_1-\omega_3)},
\end{aligned}
\label{eq:N4_ratio}
\end{equation}
where $C_{14}=C_{13}C_{34}$ follows 
from $C_{\ell\ell'}=d_1(\ell)/d_1(\ell')$ in Eq.~\eqref{eq:sync_ratio}, since the intermediate amplitude $d_1(3)$ cancels. The matching condition in Eq.~\eqref{eq:N4_condition} depends on the segment frequencies and the coupling strength $J_3$. Once the shared energy is chosen, the dimer eigenvector fixes $C_{34}$, while the couplings $J_1$ and $J_2$ control the relative weights across the damped site. Varying their ratio adjusts $C_{13}$ and $C_{14}$ without changing $C_{34}$ or the dark-mode count. The frequency $\omega_2$ of the damped qubit is unrestricted.

For the initial state $|\psi_0\rangle=|+\rangle_1|0\rangle_2|0\rangle_3|0\rangle_4\equiv \ket{+000}$, the asymptotic expectation values can be obtained explicitly for any parameters satisfying Eq.~\eqref{eq:N4_condition}.
The normalization factor of the dark mode in
Eq.~\eqref{eq:N4_mode_general} is
\begin{equation}
\mathcal{N}^2
=J_2^2+J_1^2\left[
1+\frac{(E_1-\omega_3)^2}{J_3^2}
\right].
\label{eq:N4_normalization}
\end{equation}
The dark-mode amplitude at the initially prepared site is $d_1(1)=J_2/\mathcal{N}$. Since $c_G=c_1=1/\sqrt{2}$ the dark-mode overlap is $a_1=J_2/(\sqrt{2}\mathcal{N})$. Substituting these and the dark-mode amplitudes into Eq.~\eqref{eq:expectation01}, we obtain
\begin{equation}
\begin{aligned}
\langle\sigma_x^{(1)}(t)\rangle_{\mathrm{DFS}}
&=\frac{J_2^2}{\mathcal{N}^2}
\cos\!\big[E_1(t_0+t)\big],\\
\langle\sigma_x^{(4)}(t)\rangle_{\mathrm{DFS}}
&=-\frac{J_1J_2(E_1-\omega_3)}
{J_3\mathcal{N}^2}
\cos\!\big[E_1(t_0+t)\big].
\end{aligned}
\label{eq:N4_expectations}
\end{equation}
The third qubit also satisfies
\begin{equation}
\langle\sigma_x^{(3)}(t)\rangle_{\mathrm{DFS}}
=\frac{J_3}{E_1-\omega_3}
\langle\sigma_x^{(4)}(t)\rangle_{\mathrm{DFS}}.
\label{eq:N4_third_qubit}
\end{equation}
All three undamped qubits exhibit persistent oscillations at the angular frequency $|E_1|$, with the amplitude ratios given in Eq.~\eqref{eq:N4_ratio}. For positive couplings, matching the lower dimer energy $E_-$ gives in-phase edge oscillations, whereas matching the upper energy $E_+$ gives antiphase oscillations. This is simply because $E_--\omega_3<0$ and $E_+-\omega_3>0$, irrespective of the detuning between the two dimer sites. Frequency tuning creates a dark mode absent in the uniform four-qubit chain, and the couplings allow its endpoint amplitude ratio to be adjusted.

\begin{figure}[t]
    \centering
    \includegraphics[width=0.45\textwidth]{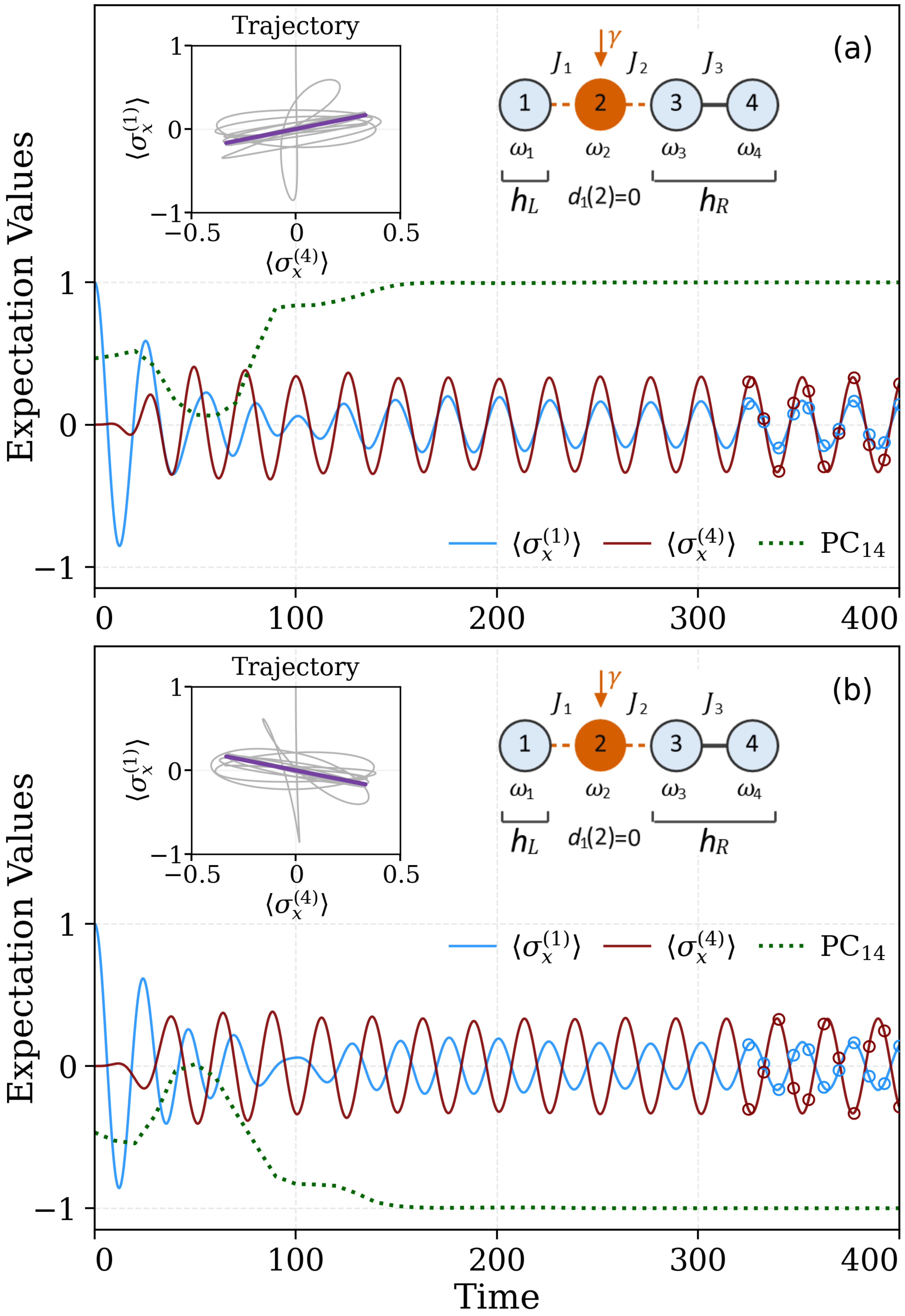}
    \caption{In-phase (a) and antiphase (b) edge synchronization in a four-qubit XX chain initially prepared in $\ket{+000}$, with couplings $J_1=J_2=J_3=0.05$ and relaxation at the second qubit at rate $\gamma=0.1$. The frequencies $(\omega_1,\omega_2,\omega_3,\omega_4)$ are fixed as $(0.25,0.30,0.35,0.275)$ and $(0.25,0.20,0.15,0.225)$ in panels (a) and (b), respectively. Blue and red curves show $\langle\sigma_x^{(1)}\rangle$ and $\langle\sigma_x^{(4)}\rangle$, while green dotted curves display their Pearson correlation $\mathrm{PC}_{14}$. Circles show the asymptotic predictions of Eq.~\eqref{eq:N4_expectations} for $325\leq t\leq400$. Time is measured from the initial preparation. Trajectory insets show the evolution for $0\leq t\leq400$ (gray) and $900\leq t\leq1000$ (purple), approaching straight lines with slopes $C_{14}=1/2$ in (a) and $C_{14}=-1/2$ in (b). Schematics indicate the dark-mode node and the two segments, with dashed bonds representing the joining couplings. $\mathrm{PC}_{14}$ is evaluated for sliding windows of $400$ samples with an overlap of $300$ samples and a time step of $0.1$.}
    \label{fig:Fig1}
\end{figure}

Figure~\ref{fig:Fig1} shows the local expectation values obtained by numerically solving the full Lindblad equation, Eq.~\eqref{eq:lindblad}, for the parameters specified in the caption. Both cases have the same dark-mode energy $E_1=0.25$, but this energy corresponds to different dimer eigenstates. In panel (a), the dimer frequencies $(\omega_3,\omega_4)=(0.35,0.275)$ lead to $E_-=0.25$ and $E_+=0.375$, so the first qubit is resonant with the lower dimer energy. On the other hand, in panel (b), the frequencies $(\omega_3,\omega_4)=(0.15,0.225)$ give $E_-=0.125$ and $E_+=0.25$, so the same first-qubit frequency matches the upper dimer energy. Hence, the edge amplitudes have the same sign in (a) and opposite signs in (b), yielding $C_{14}=1/2$ and $C_{14}=-1/2$, respectively. The common dark-mode energy fixes the same oscillation frequency, while the different spatial profiles determine whether the edge oscillations are in phase or in antiphase. During the transient, the expectation values contain decaying contributions and are not proportional. As these contributions fade, the numerical curves approach the stable oscillations predicted by Eq.~\eqref{eq:N4_expectations}. To quantify the onset of synchronization, we use the Pearson coefficient between the local expectation values at sites $\ell$ and $\ell'$,
\begin{equation}
\mathrm{PC}_{\ell\ell'}
=\frac{\operatorname{Cov}(x,y)}
{\sqrt{\operatorname{Var}(x)\operatorname{Var}(y)}},
\label{eq:pearson}
\end{equation}
where $x=\langle\sigma_x^{(\ell)}(t)\rangle$ and $y=\langle\sigma_x^{(\ell')}(t)\rangle$. Here, $\operatorname{Cov}$ and $\operatorname{Var}$ denote the covariance and variance evaluated within each sliding time window. The coefficient ranges from $-1$ to $+1$. For proportional oscillations, $\mathrm{PC}_{\ell\ell'}=+1$ corresponds to in-phase synchronization, whereas $\mathrm{PC}_{\ell\ell'}=-1$
corresponds to antiphase synchronization, independently of their amplitude ratio. The trajectory insets provide a complementary view of the emergence of synchronization by displaying the expectation values for the edge qubits. At late times, the curved transient trajectories approach straight-line segments through the origin, expressing the proportionality of the two expectation values. The slope is precisely $C_{14}$. Its sign distinguishes in-phase from antiphase oscillations, while its magnitude gives their amplitude ratio. The slope values $\pm1/2$ show that qubit 1 oscillates with half the amplitude of qubit 4 in both cases.

\subsection{Five-Qubit Chain}

The five-qubit chain with relaxation at the middle site, $M=\{3\}$, is the most basic qubit chain that can support two one-excitation dark modes. It provides a simple setting for distinguishing synchronization associated with a restricted class of initial states from generic synchronization. Here, the two undamped segments are the dimers $L=\{1,2\}$ and $R=\{4,5\}$, so their spectra might have zero, one, or two shared eigenvalues. We first consider a uniform chain and then show how frequency tuning can select a single shared energy without changing the damping configuration, leading to generic synchronization.

For $\omega_i=\omega$ and $J_i=J$, with $\omega>J>0$, both dimers have the energies $E_1=\omega+J$ and $E_2=\omega-J$. Consequently, $r=2$, with dark modes
\begin{equation}
\begin{aligned}
|D_1\rangle&=\frac{|1\rangle+|2\rangle-|4\rangle-|5\rangle}{2},\\
|D_2\rangle&=\frac{|1\rangle-|2\rangle+|4\rangle-|5\rangle}{2}.
\end{aligned}
\label{eq:N5_uniform_modes}
\end{equation}
The full DFS also contains the zero-excitation state $|G\rangle$ and the two-excitation dark state $|D_{12}\rangle$ with energy $2\omega$, whose determinant amplitudes are $d_1(i)d_2(j)-d_1(j)d_2(i)$.
Substituting the corresponding transition amplitudes for the edge qubits into Eq.~\eqref{eq:formula} gives
\begin{equation}
\begin{aligned}
\langle\sigma_x^{(1)}(t)\rangle_{\mathrm{DFS}}
&=\operatorname{Re}\!\left[A_1e^{-iE_1t}+A_2e^{-iE_2t}\right],\\
\langle\sigma_x^{(5)}(t)\rangle_{\mathrm{DFS}}
&=\operatorname{Re}\!\left[B_1e^{-iE_1t}+B_2e^{-iE_2t}\right],
\end{aligned}
\label{eq:N5_general_expectations}
\end{equation}
where the coefficients are given by
\begin{equation}
\begin{aligned}
A_1=\rho^\infty_{D_1G}+\rho^\infty_{D_{12}D_2}, \quad
B_1=-\rho^\infty_{D_1G}+\rho^\infty_{D_{12}D_2},\\
A_2=\rho^\infty_{D_2G}-\rho^\infty_{D_{12}D_1}, \quad
B_2=-\rho^\infty_{D_2G}-\rho^\infty_{D_{12}D_1}.
\end{aligned}
\label{eq:N5_coherence_coefficients}
\end{equation}
It is straightforward to see that, at each frequency, the coherence between the zero-excitation dark state and a one-excitation dark mode enters the edge expectation values with opposite signs. By contrast, the coherence between a one-excitation dark mode and the two-excitation dark state enters with the same sign. Their relative contributions depend on the preparation and the dissipative transient, so the two expectation values need not remain proportional. This is the obstruction to generic synchronization in the present four-dimensional DFS.

For an initial state containing at most one  excitation, however, the evolution cannot populate $|D_{12}\rangle$. The corresponding coherences in Eq.~\eqref{eq:N5_coherence_coefficients} vanish, giving $B_1=-A_1$ and $B_2=-A_2$, and hence
\begin{equation}
\langle\sigma_x^{(5)}(t)\rangle_{\mathrm{DFS}}
=-\langle\sigma_x^{(1)}(t)\rangle_{\mathrm{DFS}}.
\label{eq:N5_restricted_sync}
\end{equation}
That is, whenever persistent oscillations survive, the edge qubits are stably synchronized in antiphase for this class of initial states, despite the presence of two one-excitation dark modes. If both $A_1$ and $A_2$ are nonzero, the synchronized oscillations contain the two frequencies $E_1$ and $E_2$. We emphasize that stable synchronization here depends on the initial preparation and is thus not generic. In particular, let us consider the initial state $|\psi_0\rangle=|+\rangle_1|0\rangle_2|0\rangle_3 |0\rangle_4|0\rangle_5\equiv \ket{+0000}$. The vacuum amplitude is $c_G=1/\sqrt{2}$, and since both dark modes have amplitude $1/2$ at the first site, the initial dark-mode overlaps are $a_1=a_2=1/(2\sqrt{2})$. Equation~ \eqref{eq:expectation01} gives
\begin{equation}
\begin{aligned}
\langle\sigma_x^{(1)}(t)\rangle_{\mathrm{DFS}}
&=\frac12\cos\!\big[\omega(t_0+t)\big]
\cos\!\big[J(t_0+t)\big],\\
\langle\sigma_x^{(5)}(t)\rangle_{\mathrm{DFS}}
&=-\frac12\cos\!\big[\omega(t_0+t)\big]
\cos\!\big[J(t_0+t)\big].
\end{aligned}
\label{eq:N5_uniform_expectations}
\end{equation}
Here both dark modes contribute at angular frequencies $E_1=\omega+J$ and $E_2=\omega-J$, while the two edge expectation values remain exactly opposite.

Let us next consider an inhomogeneous chain with the same damping configuration and examine how tuning the qubit frequencies can make synchronization generic. We keep the internal couplings of the two dimers equal, i.e., $J_1=J_4=J$, but allow the joining couplings $J_2$ and $J_3$ to differ. Let the frequencies of the undamped qubits be
\begin{equation}
\omega_1=\omega_2=\omega, \qquad
\omega_4=\omega_5=\omega+\Delta,
\label{eq:N5_detuning}
\end{equation}
while the frequency $\omega_3$ of the damped qubit remains unrestricted. In this case, the two segment spectra are
\begin{equation}
\begin{aligned}
\operatorname{spec}(h_L)
&=\{\omega-J,\omega+J\},\\
\operatorname{spec}(h_R)
&=\{\omega+\Delta-J,\omega+\Delta+J\}.
\end{aligned}
\label{eq:N5_detuned_spectra}
\end{equation}
We can observe that at $\Delta=0$, both energies are shared and $r=2$. For $\Delta=2J$, the upper energy of $h_L$ coincides with the lower energy of $h_R$, leaving exactly one common eigenvalue. Similarly, $\Delta=-2J$ matches the lower energy of $h_L$ to the upper energy of $h_R$. Either choice gives $r=1$, while all other nonzero $\Delta$ values yield $r=0$. A suitable frequency shift can make a single dark mode survive, while an arbitrary $\Delta$ generally removes both.

We now focus on $\Delta=2J$, for which the unique shared energy is $E_1=\omega+J$. At the shared energy, the amplitudes on $L$ satisfy $d_1(2)=d_1(1)$, while those on $R$ satisfy $d_1(5)=-d_1(4)$. Also, the scalar relation at the damped site reads $J_2d_1(2)+J_3d_1(4)=0$, which fixes the relative weights of the two segments. The dark mode is
\begin{equation}
|D_1\rangle=
\frac{J_3}{\mathcal N}(|1\rangle+|2\rangle) -\frac{J_2}{\mathcal N}(|4\rangle-|5\rangle),
\label{eq:N5_tuned_mode}
\end{equation}
where $\mathcal N^2=2(J_2^2+J_3^2)$. 
The DFS contains only $|G\rangle$ and $|D_1\rangle$. Because $E_1>0$ under the assumption $\omega>J>0$, any
preparation producing persistent oscillations leads to the same proportionality between the local expectation values. This means that synchronization is generic without any restriction to initial states having at most one excitation. In particular, the edge locking constant is
\begin{equation}
C_{15}
=\frac{d_1(1)}{d_1(5)}
=\frac{J_3}{J_2}.
\label{eq:N5_tuned_ratio}
\end{equation}
For positive joining couplings, the edge oscillations are in phase. Their amplitude ratio need not be unity and can be adjusted through $J_3/J_2$ without changing the spectral matching condition or the dark-mode energy. For the same initial preparation $\ket{+0000}$, now evolving under the inhomogeneous Hamiltonian, the vacuum amplitude is $c_G=1/\sqrt{2}$ and the dark-mode overlap is $a_1=J_3/(\sqrt{2}\mathcal N)$. Substituting these coefficients and the site amplitudes of Eq.~\eqref{eq:N5_tuned_mode} into Eq.~\eqref{eq:expectation01} gives
\begin{equation}
\begin{aligned}
\langle\sigma_x^{(1)}(t)\rangle_{\mathrm{DFS}}
&=\frac{J_3^2}{\mathcal N^2}\cos[E_1(t_0+t)],\\
\langle\sigma_x^{(5)}(t)\rangle_{\mathrm{DFS}}
&=\frac{J_2J_3}{\mathcal N^2}\cos[E_1(t_0+t)].
\end{aligned}
\label{eq:N5_tuned_expectations}
\end{equation}
The second qubit has the same asymptotic expectation value as the first, the fourth has the opposite value to the fifth, and the damped qubit has zero asymptotic expectation value. The two neighbors of the damped site lock with $C_{24}=-J_3/J_2$. For the chosen initial state, the edge locking constant changes from $-1$ in the uniform chain to $J_3/J_2$ in the tuned chain, and the latter value holds for every preparation producing persistent oscillations.

\begin{figure}[t]
    \centering
    \includegraphics[width=0.45\textwidth]{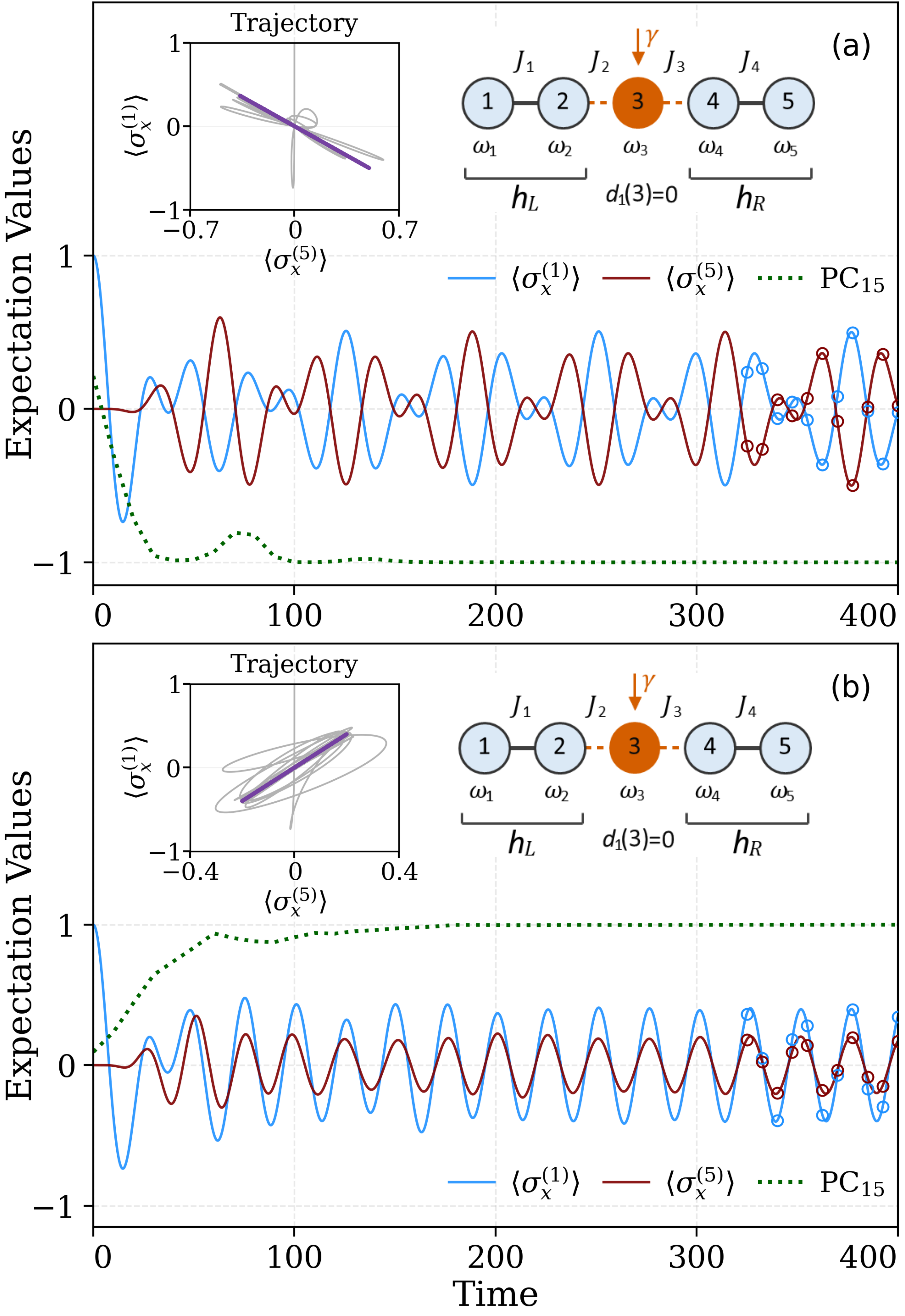}
    \caption{
    Preparation-dependent antiphase (a) and generic in-phase (b) edge synchronization in a five-qubit XX chain initially prepared in $\ket{+0000}$, with relaxation at the middle qubit at rate $\gamma=0.1$. In (a), $\omega_i=0.20$ and $J_i=0.05$. In (b), the frequencies $(\omega_1,\omega_2,\omega_3,\omega_4,\omega_5)$ are $(0.20,0.20,0.25,0.30,0.30)$, with couplings $J_1=J_2=J_4=0.05$ and $J_3=0.10$. Blue and red curves show $\langle\sigma_x^{(1)}\rangle$ and $\langle\sigma_x^{(5)}\rangle$, respectively, whereas green dotted curves display their Pearson correlation $\mathrm{PC}_{15}$. Circles show the asymptotic predictions of Eq.~\eqref{eq:N5_uniform_expectations} in (a) and Eq.~\eqref{eq:N5_tuned_expectations} in (b) for $325\leq t\leq400$. Time is measured from the initial preparation. Trajectory insets show the evolution for $0\leq t\leq400$ (gray) and $900\leq t\leq1000$ (purple), approaching straight lines with slopes $C_{15}=-1$ and $2$, respectively. Schematics indicate the dark-mode node and the two segments, with dashed bonds representing the joining couplings. $\mathrm{PC}_{15}$ is evaluated for sliding windows of $400$ samples with an overlap of $300$ samples and a time step of $0.1$.}
    \label{fig:Fig2}
\end{figure}

Figure~\ref{fig:Fig2} illustrates these two synchronization regimes through numerical solutions of the Lindblad equation for the same initial preparation. In panel (a), both surviving frequencies $E_1=0.25$ and $E_2=0.15$ contribute to the persistent oscillations, while the edge expectation values become opposite after the transient. In panel (b), frequency tuning leaves only the shared energy $E_1=0.25$, producing in-phase oscillations with amplitudes $0.4$ and $0.2$. The Pearson coefficients approach $-1$ and $+1$, respectively. The trajectory insets show how the transient motion approaches a straight line, whose slope gives the locking constant $C_{15}=-1$ in (a) and $C_{15}=2$ in (b). The asymptotic numerical results are in excellent agreement with the analytical predictions shown by the open circles. Although both chains synchronize for the chosen initial state, the underlying distinction follows from their DFS structures. In the uniform chain, the antiphase relation relies on the absence of coherences involving $|D_{12}\rangle$ for this preparation. In the tuned chain, the unique one-excitation dark mode forces the same locking constant for every initial state, giving rise to generic synchronization.

\subsection{Odd N-qubit Chains with Noise on Even Sites}
\label{sec:odd_even}

Let us first consider a chain of odd length $N=2K+1$, with $K\geq1$, in which every even site is damped, that is, $M=\{2,4,\ldots,N-1\}$. The damped sites divide the chain into $K+1$ segments, each consisting of a single odd site. Each segment contributes only its local qubit frequency to the spectral matching condition. The segments share an eigenvalue if and only if all qubits at odd sites of the chain are resonant,
\begin{equation}
\omega_1=\omega_3=\cdots=\omega_N=E_1.
\label{eq:odd_resonance}
\end{equation}
Since each segment contains only one energy, Eq.~\eqref{eq:dark_count_multiple} gives exactly one dark mode when this condition holds, whereas otherwise $r=0$. In fact, Eq.~\eqref{eq:odd_resonance} generalizes the three-qubit resonance condition of Eq.~\eqref{eq:N3_condition}. It involves neither the frequencies of the damped qubits nor the couplings. The dark mode vanishes at every even site. As a consequence, at a damped site $m$, the eigenvalue equation reduces to the scalar relation of Eq.~\eqref{eq:scalar},
\begin{equation*}
J_{m-1}d_1(m-1)+J_m d_1(m+1)=0.
\end{equation*}
Because all couplings are nonzero, a zero amplitude at any odd site would force the amplitudes at all other odd sites to vanish as well. Accordingly, every odd site carries a nonzero dark-mode amplitude. The DFS is spanned by $|G\rangle$ and $|D_1\rangle$, so for $E_1\neq0$, all odd-site qubits are generically synchronized. The damped even sites have vanishing asymptotic expectation values of $\sigma_x$.

The scalar relation fixes the locking constant between the two neighbors of each damped site as
\begin{equation}
C_{m-1,m+1}
=\frac{d_1(m-1)}{d_1(m+1)}
=-\frac{J_m}{J_{m-1}}.
\label{eq:odd_neighbor}
\end{equation}
Their relative oscillation amplitude depends only on the two couplings attached to the intervening damped qubit. Due to the fact that the locking constants are ratios of dark-mode amplitudes, they multiply along the chain, i.e., $C_{\ell\ell''}=C_{\ell\ell'}C_{\ell'\ell''}$. The locking constant between any two odd sites is the product of the factors in Eq.~\eqref{eq:odd_neighbor} associated with the damped sites between them. For the endpoints of the chain, this gives
\begin{equation}
C_{1N}
=(-1)^K
\frac{J_2J_4\cdots J_{N-1}}
     {J_1J_3\cdots J_{N-2}}.
\label{eq:odd_edge}
\end{equation}
Each damped site contributes one minus sign and one coupling ratio. These constants also determine the entire spatial profile through $d_1(\ell)=d_1(1)/C_{1\ell}$ at every odd site, with $C_{11}=1$. Normalization fixes the magnitude of the first-site amplitude as
\begin{equation}
\frac{1}{d_1(1)^2}
=1+\frac{1}{C_{13}^2}+\cdots+\frac{1}{C_{1N}^2}.
\label{eq:odd_norm}
\end{equation}
The sum contains one contribution from each odd site. Together, Eqs.~\eqref{eq:odd_neighbor} and \eqref{eq:odd_norm} determine the normalized dark mode up to an irrelevant overall sign. For $N=3$, Eq.~\eqref{eq:odd_neighbor} reproduces the locking constant in Eq.~\eqref{eq:N3_ratio}, and normalization recovers
the dark mode in Eq.~\eqref{eq:N3_mode}. For positive couplings, every constant in Eq.~\eqref{eq:odd_neighbor} is negative, so neighboring odd sites oscillate in antiphase. The endpoints oscillate in phase when $K$ is even, corresponding to $N=5,9,13,\ldots$, and in antiphase when $K$ is odd, corresponding to $N=3,7,11,\ldots$. The magnitudes of the locking constants are set by the coupling ratios and can take any positive value. For equal couplings, all odd-site amplitudes have an identical magnitude, $1/\sqrt{K+1}$, and $C_{1N}=(-1)^K$. If all qubit frequencies are also equal, the chain is fully uniform, and its dark-mode count agrees with $r=1$ obtained from Eq.~\eqref{eq:uniform_gcd}.

For an initial state of the form $\ket{+0\cdots0}$, the vacuum amplitude and dark-mode overlap read $c_G=1/\sqrt{2}$ and $a_1=d_1(1)/\sqrt{2}$. Equation~\eqref{eq:expectation01} gives, at every odd site $\ell$,
\begin{equation}
\langle\sigma_x^{(\ell)}(t)\rangle_{\mathrm{DFS}}
=\frac{d_1(1)^2}{C_{1\ell}}
\cos\!\big[E_1(t_0+t)\big].
\label{eq:odd_expectations}
\end{equation}
The oscillations have angular frequency $|E_1|$, and their relative signs and amplitudes are obtained from the locking constants.  As a concrete example, let us next consider $N=5$ with $M=\{2,4\}$, under the resonance condition $\omega_1=\omega_3=\omega_5=E_1$. Equation~\eqref{eq:odd_neighbor} gives $C_{13}=-J_2/J_1$ and $C_{35}=-J_4/J_3$. Their product determines the endpoint ratio, $C_{15}=J_2J_4/(J_1J_3)$, and the dark mode is
\begin{equation}
|D_1\rangle
=\frac{J_2J_4|1\rangle-J_1J_4|3\rangle+J_1J_3|5\rangle}
{\sqrt{J_2^2J_4^2+J_1^2J_4^2+J_1^2J_3^2}}.
\label{eq:odd_N5_mode}
\end{equation}
For $J_1=J_3=0.05$ and $J_2=J_4=0.1$, the local locking constants are $C_{13}=C_{35}=-2$, giving $C_{15}=4$. Equation~\eqref{eq:odd_norm} yields $d_1(1)^2=1/(1+1/4+1/16)=16/21$. For the initial state $\ket{+0000}$, the three asymptotic expectation values therefore share the factor $\cos[E_1(t_0+t)]$, with respective coefficients $16/21$, $-8/21$, and $4/21$ at sites 1, 3, and 5. That is, the endpoints oscillate in phase, while the central qubit oscillates in antiphase with both.

The couplings in the above example alternate between two values, while the coefficients change by a factor of $-1/2$ from one odd site to the next. Extending this alternation to arbitrary odd length gives a dimerized chain, where $J_{2a-1}=v$ and $J_{2a}=w$ with $v,w>0$. Under the Jordan-Wigner transformation, the XX chain maps onto noninteracting fermions with on-site energies $\omega_i$ and hopping amplitudes $J_i$, whose single-particle Hamiltonian is $H^{(1)}$. For alternating hoppings and all qubits resonant at $\omega$, these fermions realize the Su-Schrieffer-Heeger (SSH) model~\cite{Su1979}, so the dimerized XX chain is its spin counterpart and its one-excitation modes are the single-particle SSH modes. With every even site damped as before, all neighbor locking constants in Eq.~\eqref{eq:odd_neighbor} equal $-w/v$, and the dark-mode amplitudes form a geometric sequence,
\begin{equation}
d_1(2k+1)=\left(-\frac{v}{w}\right)^{k}d_1(1),
\quad
d_1(1)^2=\frac{1-(v/w)^2}{1-(v/w)^{N+1}},
\label{eq:ssh_profile}
\end{equation}
where the normalization constant is obtained by summing the geometric series in Eq.~\eqref{eq:odd_norm} for $v\neq w$. For $N=5$ and $v/w=1/2$, this gives $d_1(1)^2=16/21$, as in the example above. For $v<w$, the strong bonds bind the qubit pairs $(2,3),(4,5),\ldots,(N-1,N)$ into dimers and leave the first qubit unpaired. The dark mode is the edge state of the SSH chain, which has the energy $E_1=\omega$ and decays by a factor $v/w$ per unit cell away from the first qubit. Its existence does not depend on the dimerization, since Eqs.~\eqref{eq:odd_neighbor} to \eqref{eq:odd_norm} hold for any couplings, and the dimerization only localizes it at one edge. All odd-site qubits are synchronized, with oscillation amplitudes that decrease along the chain, and the locking constant $C_{1N}=(-w/v)^K$ grows exponentially in magnitude with the chain length. For the initial state $\ket{+0\cdots0}$, Eq.~\eqref{eq:odd_expectations} gives the amplitude $d_1(1)^2(-v/w)^k$ at site $2k+1$. For $v/w=1/2$, for instance, the last qubit oscillates in phase with the first, with $1/1024$ of its amplitude for $N=21$. On the other hand, for $v>w$, the edge state is localized near the last qubit instead, while $v=w$ recovers the equal amplitudes of the uniform chain.

The same mode remains dark if only a subset of the even sites is damped, since it vanishes at every even site. With fewer damped sites, however, the segments become longer and may share additional eigenvalues, so the mode need not be the only dark mode. In any case, a single-site segment still limits the chain to at most one dark mode.

\section{Entanglement between Synchronized Qubits}
\label{sec:entanglement}

Synchronization concerns the time dependence of local observables, whereas entanglement characterizes the joint state of different qubits. We now examine their relationship in the asymptotic dynamics. When $r=1$, the DFS is spanned by the vacuum $|G\rangle$ and the unique dark mode $|D_1\rangle$. Its density operator can be written as
\begin{equation}
\begin{aligned}
\rho_{\mathrm{DFS}}(t)
={}&(1-p)|G\rangle\langle G|
+p|D_1\rangle\langle D_1|\\
&+z(t)|D_1\rangle\langle G|
+z(t)^*|G\rangle\langle D_1|,
\end{aligned}
\label{eq:entanglement_DFS}
\end{equation}
where $p=\rho^\infty_{D_1D_1}$ is the surviving dark-mode population and $z(t)=\rho^\infty_{D_1G}e^{-iE_1t}$ is its coherence with the vacuum. The population remains constant during the asymptotic evolution, while the coherence acquires the phase determined by $E_1$. To determine the entanglement between qubits $j$ and $k$, we trace out the remaining qubits. Writing $d_j=d_1(j)$ and $d_k=d_1(k)$, the reduced density operator in the basis $\{|00\rangle,|01\rangle,|10\rangle,|11\rangle\}$ is
\begin{equation}
\rho_{\mathrm{DFS}}^{(jk)}(t)=
\begin{pmatrix}
1-p(d_j^2+d_k^2) & d_kz^* & d_jz^* & 0\\
d_kz & pd_k^2 & pd_jd_k & 0\\
d_jz & pd_jd_k & pd_j^2 & 0\\
0 & 0 & 0 & 0
\end{pmatrix},
\label{eq:entanglement_reduced}
\end{equation}
where $z=z(t)$. The $|00\rangle$ population includes both the global vacuum population $1-p$ and the probability $p(1-d_j^2-d_k^2)$ that the dark-mode excitation lies outside the selected pair. The $|11\rangle$ row and column vanish because every state in this DFS contains at most one excitation.

We quantify the pairwise entanglement using the concurrence, $\mathcal C(\rho)
=\max\left\{0,\sqrt{\lambda_1}-\sqrt{\lambda_2} -\sqrt{\lambda_3}-\sqrt{\lambda_4}\right\}$, where the $\lambda_i$ are the eigenvalues of $\rho\widetilde\rho$ in decreasing order, with $\widetilde\rho=(\sigma_y\otimes\sigma_y)\rho^*(\sigma_y\otimes\sigma_y)$~\cite{Wootters1998}. For a density operator with no support on $|11\rangle$, the two potentially nonzero eigenvalues are
\begin{equation}
\lambda_\pm
=\left(
\sqrt{\rho_{01,01}\rho_{10,10}}
\pm|\rho_{10,01}|
\right)^2.
\label{eq:concurrence_eigenvalues}
\end{equation}
Positivity requires $\rho_{01,01}\rho_{10,10}\geq|\rho_{10,01}|^2$, which implies that $\mathcal C=2|\rho_{10,01}|$, independently of the coherences with $|00\rangle$. In Eq.~\eqref{eq:entanglement_reduced}, the one-excitation block has rank one if $p(d_j^2+d_k^2)>0$, and $\lambda_-=0$. Consequently, the asymptotic concurrence becomes
\begin{equation}
\mathcal C^\infty_{jk}
=2p\,|d_1(j)d_1(k)|,
\label{eq:concurrence_r1}
\end{equation}
which is obviously constant in time, even when the local expectation values oscillate persistently.

This expression establishes a direct implication for the generically synchronized regime. Persistent oscillations require $\rho^\infty_{D_1G}\neq0$, while the positivity of the DFS density operator imposes
\begin{equation}
|\rho^\infty_{D_1G}|^2\leq p(1-p).
\label{eq:entanglement_positivity}
\end{equation}
A nonzero coherence between the vacuum and the dark mode consequently requires $p>0$. Every pair of qubits participating in the synchronized oscillations has nonzero dark-mode amplitudes and hence a strictly positive, constant concurrence. The converse does not hold~\cite{Roulet2018b}, as also observed for the uniform chain in Ref.~\cite{Cakmak2026}. A state with $p>0$ but $\rho^\infty_{D_1G}=0$ is entangled on every pair within the dark-mode support, although all local expectation values of $\sigma_x$ vanish. Synchronization depends on coherence between the vacuum and the dark mode, whereas the concurrence depends on the dark-mode population. The spatial profile determines both the locking constants and the distribution of pairwise entanglement, but the magnitude of the entanglement also depends on $p$, which can retain information about the initial preparation and the dissipative transient dynamics.

For a pure initial state containing at most one excitation, relaxation does not feed the dark mode, and its surviving population is  $p=|\langle D_1|\psi_0\rangle|^2$. In particular, the preparation $\ket{+0\cdots0}$ gives $p=d_1(1)^2/2$, and hence
\begin{equation}
\mathcal C^\infty_{jk}
=d_1(1)^2\,|d_1(j)d_1(k)|.
\label{eq:concurrence_initial01}
\end{equation}
For a uniform chain with a unique dark mode, the endpoint amplitudes satisfy $d_1(1)^2=d_1(N)^2=2/(N+1)$. The endpoint concurrence for this preparation becomes
\begin{equation}
\mathcal C^\infty_{1N}=\frac{4}{(N+1)^2},
\label{eq:concurrence_uniform}
\end{equation}
recovering the uniform-chain result of Ref.~\cite{Cakmak2026}. When several dark modes exist, constant concurrence is no longer guaranteed. For initial states containing at most one excitation, the reduced pair still has no $|11\rangle$ component, so its concurrence remains twice the magnitude of its one-excitation coherence, i.e.,
\begin{equation}
\begin{aligned}
\mathcal C^{\mathrm{DFS}}_{jk}(t)
=2\Bigg|
\sum_{\alpha,\beta=1}^{r}
\rho^\infty_{D_\alpha D_\beta}
d_\alpha(j)d_\beta(k)
\times e^{-i(E_\alpha-E_\beta)t}
\Bigg|.
\end{aligned}
\label{eq:concurrence_multiple}
\end{equation}
Coherences between distinct dark modes can produce time-dependent entanglement due to energy differences. 

The uniform five-qubit chain with damping at the middle site provides an explicit example of synchronization accompanied by varying entanglement. For the preparation $\ket{+0000}$, the two edge expectation values remain opposite throughout the asymptotic evolution, while their concurrence is
\begin{equation}
\mathcal C^{\mathrm{DFS}}_{15}(t)
=\frac14\cos^2[J(t_0+t)].
\label{eq:concurrence_N5}
\end{equation}
Thus, the state-dependent synchronization supported by two dark modes can coexist with oscillating concurrence. Conversely, a coherent superposition confined to the one-excitation dark subspace has vanishing local expectation values of $\sigma_x$, yet its pairwise concurrence can oscillate. Time-independent concurrence is guaranteed by the $r=1$ DFS structure, but synchronization and entanglement dynamics are distinct properties of the surviving state.

\section{Conclusion}
\label{sec:conc}

In this work, we studied stable synchronization in XX chains with arbitrary site frequencies and exchange couplings under local and multi-local amplitude damping. We characterized the full DFS through its one-excitation dark modes and, for initial states with at most one excitation, obtained the asymptotic local dynamics without solving the dissipative transient. We showed that the dark-mode energies are the eigenvalues shared by all undamped segments separated by the damped sites, and that each shared eigenvalue gives exactly one dark mode. Therefore, the number of dark modes is the degree of the greatest common divisor of the segment characteristic polynomials, which recovers the integer rule of the uniform chain~\cite{Cakmak2026} as a special case. Generic stable synchronization occurs when the segments share a single eigenvalue of nonzero energy. The spectral matching criterion also separates the roles of the Hamiltonian parameters. The frequencies and couplings within the segments determine whether dark modes exist and fix their energies, while the frequencies of the damped qubits play no role. The couplings that join the segments to the damped sites only shape the spatial profile of a dark mode, and thereby set the amplitude ratios and relative signs of the synchronized oscillations. Since the locking constants are independent of the damping rates, the synchronized pattern is fixed by the Hamiltonian and the damping configuration, while the rates determine how it is approached. 

Through analytical examples, we showed that suitable detunings can create synchronization that is forbidden in the corresponding uniform chain and turn preparation-dependent locking into generic synchronization. For odd chains damped at every even site, resonance of the odd-site qubits alone yields a single dark mode for any choice of couplings, with the coupling profile determining how the surviving oscillations are distributed along the chain. We also derived the asymptotic concurrence for any pair of qubits. In the single-dark-mode regime, the concurrence is constant in time, and every pair participating in synchronized oscillations remains entangled, with a value set by the dark-mode profile and its population. Since the local oscillations additionally require coherence between the vacuum and the dark mode, entanglement can persist without synchronization. When several dark modes are present, preparation-dependent synchronization can coexist with a time-dependent concurrence.

Our DFS construction relies on the free-fermion mapping and excitation-number conservation, allowing multi-excitation dark states to be built from one-excitation dark modes. In case of the open nearest-neighbor chains, the tridiagonal structure of the one-excitation Hamiltonian further reduces the dark-mode problem to spectral matching between the undamped segments. Hence, our approach can be applied to other excitation-preserving chains having a free-fermion mapping, such as open XX chains with a nearest-neighbor Dzyaloshinskii-Moriya interaction along the $z$ axis~\cite{Dzyaloshinsky1958,Moriya1960,Derzhko2006}. The engineered couplings used for perfect state transfer~\cite{Christandl2004,Kay2010} are a special case of the chains studied here, so the spectral matching condition directly determines which damping configurations leave them with dark modes. When the segment spectra match only approximately, the dark mode becomes slowly decaying, and synchronization can persist as a long-lived transient~\cite{Giorgi2019}.  Finally, our results can be tested on platforms where qubit frequencies can be tuned and dissipation can be engineered on selected sites, such as superconducting circuits~\cite{Tao2025} and trapped ions~\cite{Li2025,Wachtler2025}.

\acknowledgements
B.Ç., K.S., E.Y.A. and G.K. are supported by the Scientific and Technological Research Council of Türkiye (TUBITAK) under Grant No. 125F434. B.Ç. is partially supported by the Farmingdale State College Provost's Office Summer Research Scholarship. G.K. is supported by the Scientific and Technological Research Council of Türkiye (TUBITAK) through the 100th Anniversary Incentive Award. S.C. acknowledges support from the John Templeton Foundation Grant ID 63626 and from Taighde \'Eireann - Research Ireland under grant number 24/EPSRC/4121. This publication has emanated from research conducted with the financial support of Taighde \'Eireann – Research Ireland, under Grant No. 23/RC/12197 at Rinn Quantum. The authors used Claude Opus 5.5 (Anthropic) and Astra 6.0 (OpenAI) to assist with the literature search, language editing, numerical checks, and the polishing of analytical derivations. All incorporated suggestions were refined and verified by the authors, who take full responsibility for the content of this paper.

\bibliography{bibliography}

\end{document}